\documentclass[aps,prb,twocolumn]{revtex4-1}
\usepackage{graphicx}
\usepackage{amsmath}
\usepackage{amsfonts}
\usepackage{color}

\begin{document}

\title{Controlling Exceptional Points with Light}

\author{Ayan Banerjee}
\affiliation{Solid State and Structural Chemistry Unit, Indian Institute of Science, Bangalore 560012, India}
\author{Awadhesh Narayan}
\email{awadhesh@iisc.ac.in}
\affiliation{Solid State and Structural Chemistry Unit, Indian Institute of Science, Bangalore 560012, India}

\date{\today}

\begin{abstract}
We propose and show that application of light leads to an intriguing platform for controlling exceptional points in non-Hermitian topological systems. We demonstrate our proposal using three different non-Hermitian systems -- nodal line semimetals, semi-Dirac semimetals and Dirac semimetals -- and show that using illumination with light one can engineer the positions and stability of exceptional points. We illustrate the topological properties of these models and map out their light-driven topological phase transitions.

\end{abstract}

\maketitle
\section{Introduction} 
In the last few years, the study of topological phases in non-Hermitian systems has attracted wide interest~\cite{torres2019perspective,ghatak2019new,bergholtz2019exceptional,gong2018topological}. This rapidly burgeoning field encompasses both theoretical and experimental advancements in atomic physics~\cite{xu2017weyl,goldman2016topological}, quantum optics~\cite{ozawa2019topological,noh2018topological}, microwave cavities~\cite{poli2015selective,cao2015dielectric}, and topological laser systems~\cite{bandres2018topological,harari2018topological,feng2014single}, using controlled dissipation with incorporation of gain and loss terms in open systems. Considering the interplay between non-Hermiticity and topology, several varieties of non-Hermitian topological semimetals~\cite{kawabata2019classification} have been proposed, including knot~\cite{carlstrom2019knotted}, nodal line~\cite{wang2018non,budich2019symmetry}, nodal ring~\cite{wang2019non}, Hopf link~\cite{yang2019non}, Dirac~\cite{rui2019topology,papaj2019nodal}, Weyl\cite{xu2017weyl} and semi-Dirac~\cite{banerjee2020non}, to name just a few. Some of these have been experimentally realized recently.

Non-Hermitian topological systems show a distinct class of spectral degeneracies, known as \emph{exceptional points} (EPs), where not only eigenvalues but also eigenvectors coalesce and result in the Hamiltonian becoming non-diagonalizable~\cite{heiss2012physics,alvarez2018non}. Enclosing such an EP yields a quantized topological invariant revealing its underlying topological nature~\cite{heiss2016mathematical}.

On the other hand, there has been a growing interest in creating new topological phases of matter using light -- a field dubbed Floquet engineering~\cite{oka2009photovoltaic,lindner2011floquet,cayssol2013floquet,wang2013observation,rechtsman2013photonic,rudner2020floquet}. Application of light to low-dimensional systems, including graphene and silicene, has led to various topological phases, creating non-trivial gaps in their spectra~\cite{kitagawa2011transport,gu2011floquet,ezawa2013photoinduced,perez2014floquet}. Photon dressing and Floquet dynamics of various topological band systems including Weyl semimetals~\cite{wang2014floquet,gonzalez2016macroscopic}, nodal semimetals~\cite{narayan2016tunable,yan2016tunable,taguchi2016photovoltaic,chan2016type}, Chern insulators~\cite{inoue2010photoinduced,saha2016photoinduced}, Dirac and semi-Dirac semimetals~\cite{narayan2015floquet,gonzalez2016macroscopic} have been well studied and their rich topological properties have been intensively explored.
 
Very recently, non-Hermitian Floquet topological phases have also been studied with great interest. The study of periodically driven non-Hermitian systems opens up an avenue to explore new features of non-Hermitian topological phases. We highlight among them the study of topological edge state properties~\cite{zhang2020non,zhou2018non}, their dynamical characterization~\cite{zhou2019dynamical}, interplay with disorder~\cite{wu2020floquet} and their transport properties~\cite{hockendorf2019non,rodriguez2019topological,hockendorf2020cutting}.

Motivated by these rapid developments, in this work, we combine the two ideas and propose to use Floquet engineering to create and control tunable non-Hermitian phases. Generally, experimental setups used to study the Floquet toplogical phases are dissipative systems subject to gain and loss~\cite{alvarez2018topological,zhou2018non}. Therefore, it becomes particularly important to study non-Hermitian Floquet topological phases in search of new exciting phenomena. In this contribution, we show that using light one can control the positions and stability of EPs. We demonstrate our proposal using three examples in different dimensions, namely non-Hermitian nodal line semimetal and non-Hermitian Dirac semimetal in three dimensions and non-Hermitian semi-Dirac semimetal in two dimensions. We show that one can tune the position of the EPs, as well as create and annihilate them by using light. We use analytical as well as numerical calculations to illustrate the topological properties and map out the topological phase transitions arising from the application of light. Our hope is that these findings will motivate future theoretical and experimental investigations to engineer non-Hermitian Floquet topological phases.

\section {Nodal line semimetal in three dimensions} 
The continuum Hamiltonian describing a two-band spinless non-Hermitian nodal-line semimetals (with $\hbar=c=1$) is~\cite{yan2016tunable,wang2019non}

\begin{figure*}
\includegraphics[scale=0.3]{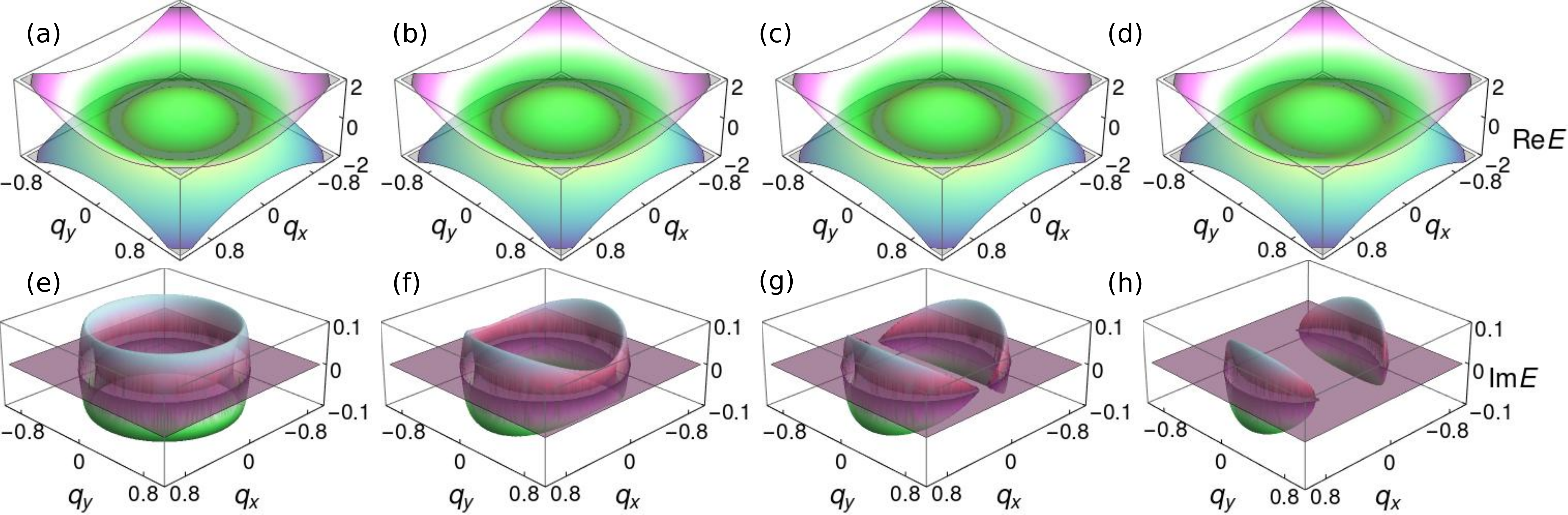}
  \caption{\textbf{Band diagrams of photon dressed non-Hermitian nodal semimetals.} (a) The real and (e) the imaginary part of the energy dispersion in the the absence of light ($a_y=a_z=0$). (b) Real and (f) imaginary part of energy spectrum with moderate laser intensity ($a_y=a_z=0.25$). We see the (c) real and (g) imaginary spectrum at critical light intensity ($a_y=a_z=0.267$), at which the rotational symmetry of nodal ring is lost along the $q_x=0$ line. Beyond this critical light intensity, (d) the real and (h) imaginary part of the spectrum are shown. The spectrum turns into two arcs. We choose the following values for the other parameters: $B=v_f=1.0$, $\eta=1$, $e=\omega=1.0$, $m=0.5$, and $\gamma=0.1$, keeping them unchanged unless otherwise specified.} \label{band_diagram}
\end{figure*}

\begin{equation}\label{model_Hamiltonian}
 H_0(\textbf q)=\epsilon_0(\textbf q)I+(m-Bq^2)\sigma_x+(v_zq_z+i\gamma)\sigma_z,
\end{equation}

where total squared momentum $q^2=q_x^2+q_y^2+q_z^2,$ $\sigma_i$ $(i=x,y,z)$ is the triad of Pauli matrices and $I$ is the identity matrix. The Pauli matrices represent the two orbital subspace. Here $v_z$ denotes the Fermi velocity along the $z$ direction, and $m$ and $B$ are parameters with the dimensions of energy and inverse energy, respectively ~\cite{yan2016tunable}. Note that the Hamiltonian is rotationally symmetric in the $q_x-q_y$ plane. Here $i\gamma\sigma_z$ is the non-Hermitian term associated with particle gain and loss between the two orbitals, with strength $\gamma$. See Appendix A for the effect of other types of non-Hermitian terms. In the absence of non-Hermitian $\gamma$ term, we get a nodal ring in the $q_z=0$ plane. The nodal ring is protected by combined inversion ($P$) and time reversal symmetries ($T$)~\cite{fang2016topological}. The non-Hermitian gain and loss term explicitly breaks the $PT$ symmetry and nodal ring splits into two exceptional rings~\cite{wang2019non}. We note here that in the case of weak coupling to a reservoir (leading to a gain and loss), the reduced density matrix of the open system obeys the Lindblad master equation. In the limit of short time dynamics, the Lindblad equation can be recast into a form with a simplified effective non-Hermitian Hamiltonian~\cite{bergholtz2019exceptional}. Furthermore, going beyond the short-time limit, it is possible to construct systems where the steady state of the Lindblad equation coincides with the long time evolution of the effective non-Hermitian Hamiltonian~\cite{diehl2011topology}. 

We obtain the energy eigenvalues as 

\begin{equation}\label{energy-eigenvalue1}
E=  \epsilon_0(\textbf q)\pm \sqrt{(m-Bq^2)^2+v^2_zq_z^2-\gamma^2+2iv_zq_z\gamma},
\end{equation}

where $\pm$ denote the conduction and valence band respectively. The energy eigenvalues are in general complex. Without loss of generality, we choose $\epsilon_0$ to be zero. In the absence of the non-Hermitian term, the conduction and valence bands touch each other and form a nodal ring in the $q_z=0$ plane for $mB>0$. The relative sign of $mB$ controls the transition between a trivial insulator phase and a nodal semimetal phase. In the presence of the non-Hermitian term, $i\gamma\sigma_z$, the original nodal ring splits into two exceptional rings (ERs) for $\gamma<m$. The band diagrams clearly demonstrating these features are presented in Fig.~\ref{band_diagram}. We note that the energy is real both inside the inner ER and outside the outer ER, while being imaginary between the two ERs.

The non-Hermitian term creates four exceptional points (EPs) along the nodal line in the $q_z=0$ plane, where the valence and conduction band touch each other. The location of the four EPs are given by $q_{\mathrm{EP}}=\pm\sqrt{(m-s\gamma)/B}$, where $s=\pm1$ in the $q_x-q_y$ plane. At the critical value of $\gamma$, i.e. $\gamma=m$, the inner EPs coincide with each other, and the inner ER becomes a point.

We now introduce an off-resonant laser beam of frequency $\omega$ polarized in the $yz$ plane on the non-Hermitian nodal semimetal. We choose the vector potential to be $\textbf{{A}}(t)=a_y\eta\cos{\omega t}\hat{y}+a_z\cos{(\omega t+\phi)}\hat{z}$, where angle $\phi$ controls the polarization of light. Here $\eta = \pm 1$ represents the left or right handedness of the incident light beam. We obtain the full time-dependent Hamiltonian considering the minimal coupling $\textbf{q}\rightarrow \textbf{q} + e\textbf{{A}}(t)$. Using Floquet formalism, the time-dependent Hamiltonian can be approximated as~\cite{kitagawa2011transport,cayssol2013floquet,rudner2020floquet}

\begin{equation}\label{effecive_floquet}
H_{\mathrm{eff}}=H + \dfrac{[H_{-1},H_{+1}]}{\omega} + O(1/\omega^2),
\end{equation}

where $H_{\pm1}=\dfrac{\omega}{2\pi}\int_{0}^{T}H(t)e^{\pm i \omega t} dt$ are Fourier coefficients of the time-dependent Hamiltonian, and $T=2\pi/\omega$ is the time period of the incident light. We note that such an approximation is valid in the high frequency domain for small laser power. The effective Hamiltonian in the presence of light is

\begin{equation}\label{effective-hamiltonian}
H_{\mathrm{eff}}=(m - B q^2) \sigma_x+(v_z q_z +i \gamma)\sigma_z + \dfrac{2 e^2 \eta B v_z a_y a_z q_y \sin{\phi}}{\omega}\sigma_y.
\end{equation}

\begin{figure*}
\includegraphics[scale=0.3]{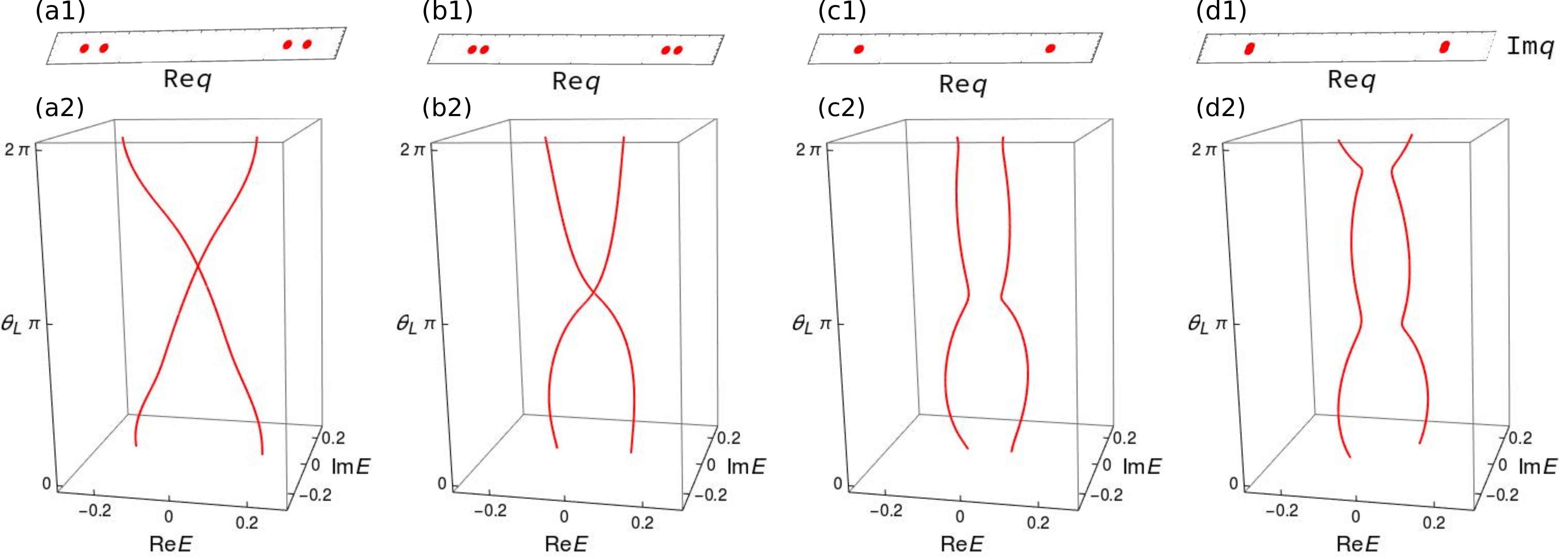}
  \caption{\textbf{Locations of exceptional points in nodal line semimetals in the presence of light illustrating the topological phase transitions.} The trajectory of the two complex eigenvalues, when the contour parameterized by $\theta_L \in [0,2\pi)$ enclosing the EP with light intensity (a2) $a_y=a_z=a=0.0$, (b2) $a_y=a_z=a=0.37$, (c2) $a_y=a_z=a=0.38$, (d2) $a_y=a_z=a=0.4$. The position of EPs in (a1), (b1), (c1), and (d1) with same light intensity as in the vorticity diagrams. Note that the four EPs move identically for $\gamma < m$. At the critical laser intensity [(c1) and (c2)], the EPs annihilate pairwise and the two complex energy bands yield a trivial vorticity. Beyond this critical laser intensity [(d1) and (d2)], the imaginary coordinates of EPs indicate that they do not exist anymore, as also confirmed by zero vorticity value. The upper panels contain four EPs with alternate $\pm 1/2$ vorticities from left to right respectively. We have shown the quasi-energy spectrum of right-most EP as a function of increasing light intensity. Here we set $\gamma=0.2$ and $m=0.5$.} \label{Vorticity_2}
\end{figure*}

Notice that in the effective Hamiltonian, we get the photon-dressed term linear in $q_y$, explicitly breaking the rotational symmetry in the $q_x-q_y$ plane. The $\sin{\phi}$ dependence indicates that the linearly polarized light does not affect the band structure. The light-induced band structure for the non-Hermitian nodal line semimetal is shown in Fig.~\ref{band_diagram}, at different values of light intensity for circularly polarized light ($\phi=\pi/2)$. The calculated band diagram shows that in this case the light illumination does not create a gap in the spectrum, but it affects the rotational symmetry of ERs along $q_y$. As a consequence of the competition between linear and quadratic terms in $q_y$ the energy band starts losing the nodal ring symmetry along the $q_x=0$ line. With increasing light intensity, the linear term in $q_y$ dominates over the quadratic term. Eventually beyond a critical value of light intensity, the bands change topology, turning into two arcs in the $q_x-q_y$ plane.
Instead of  $yz$ plane, if we had started with the laser beam polarized in the $xz$ plane, the light-induced term would have been of the form $\dfrac{2 e^2 \eta B v_z a_x a_z q_x \sin{\phi}}{\omega}\sigma_y$. Such a term would affect the rotational symmetry along $q_x=0$. This means that with increasing light intensity, we would have obtained two arcs along $q_y=0$ line, i.e., in a direction orthogonal to the case when the laser beam is polarized in the $yz$ plane. This observation suggests that we can control the locations of the EPs along any direction in the $q_x-q_y$ plane by simply rotating the laser beam in the plane perpendicular to the plane originally containing the nodal line.

The light-induced term allows control over the position of the EPs. In order to characterize the topological properties, we adopt cylindrical polar coordinates and represent EPs and ERs in the $q_\rho-q_z$ plane. We write $q_\rho=\sqrt{q_{x}^2+q_{y}^2}$, $\theta=\tan^{-1}{q_y/q_x}$ and choose $\theta=\pi/2$. This choice gives $q_x=0$ and $q_y=q_\rho$. We see that rotational symmetry reduces one degree of freedom in the system. Upon choosing $a_x = a_y = a$ and $\phi=\pi/2$, the dispersion relation now reads $E=\pm \sqrt{(m-Bq_y^2)^2+v^2_zq_z^2-\gamma^2+2iv_zq_z\gamma+(2e^2 \eta Bv_za^2q_y/\omega)^2}$. We discover that our system now has four EPs, symmetrically situated about the nodal line, located at

\begin{equation} \label{location_EP_floquet}
q_{\mathrm{EP}}=\pm \sqrt{\dfrac{(2mB-\zeta^2)\pm\sqrt{(\zeta^2-2mB)^2-4B^2(m^2-\gamma^2)}}{2B^2}},
\end{equation}

where $\zeta=2e^2 \eta Bv_za^2/\omega$. Locations of the EPs are shown for different values of $a$ in Fig.~\ref{Vorticity_2}. We find that at a critical value of $a$, the four EPs annihilate each other and their positions become imaginary. From Eq.~\ref{location_EP_floquet}, we obtain two conditions for the EPs to disappear (i) $(\zeta^2-2mB)^2\leq 4B^2(m^2-\gamma^2)$ and (ii) $m \leq \gamma $. Since, we only consider $m \geq \gamma$, we focus on condition (i), which leads to the critical values 

\begin{equation}\label{critical_A-value}
\begin{split}
\zeta_c & = \sqrt{2mB\pm\sqrt{4B^2(m^2-\gamma^2)}} \\
a_c & = (\omega^2/4e^4B^2v_z^2)^{1/4}[2mB\pm\sqrt{4B^2(m^2-\gamma^2)}]^{1/4}.
\end{split}
\end{equation}

\begin{figure*}
\includegraphics[scale=0.3]{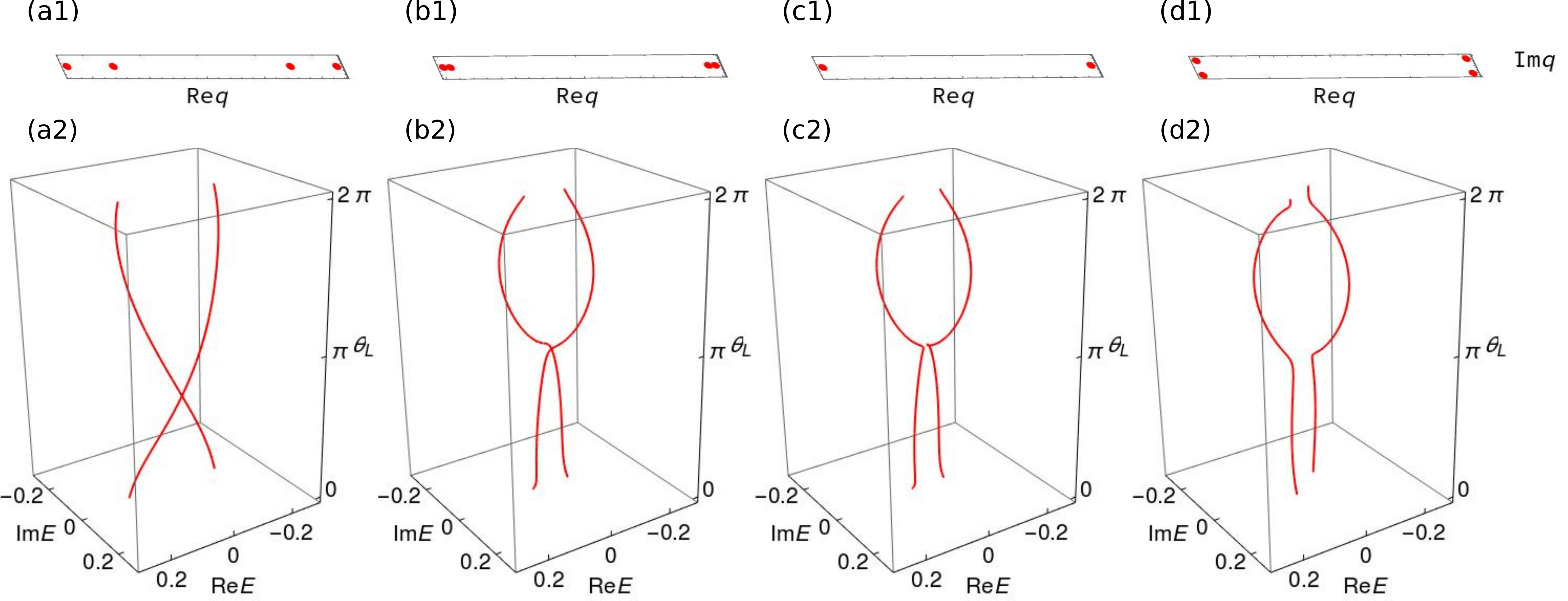}
  \caption{\textbf{Exceptional points in semi-Dirac semimetals under illumination with varying light amplitude.} The trajectory of the two complex eigenvalues, when the contour parameterized by $\theta_L \in [0,2\pi)$ enclosing the EP with light intensity (a2) $a=0.0$, (b2) $a=0.45$, (c2) $a=0.452$, (d2) $a=0.46$. The position of EPs in (a1), (b1), (c1), and (d1) with same light intensity as in the vorticity diagrams in the lower panels. At the critical laser intensity [(c1) and (c2)], the two EPs annihilate each other and the two complex energy bands yield a trivial vorticity. Beyond this critical amplitude [(d1) and (d2)], the system does not exhibit EPs. The upper panels contain four EPs with alternate $\pm 1/2$ vorticities from left to right respectively. We have shown the quasi-energy spectrum of right-most EP as a function of increasing light intensity. Here we set $\gamma=0.2$ and $\delta_0=0.5$.} \label{Vorticity_3}
\end{figure*}

We expect non-Hermitian topological phase transitions across these critical values due to the tuning by light. Next, we characterize these light induced topological phase transitions using topological invariants for non-Hermitian systems. When we encircle an EP, the associated bands get swapped in the parameter space of complex energy due to the Riemann sheet geometry of the energy bands while returning to initial states. One can define the vorticity, $\nu_{mn}$, directly associated with the complex energy dispersion for any pair of bands as~\cite{shen2018topological}

 \begin{equation}\label{vorticity}
     \nu_{mn}(\Gamma)=-\dfrac{1}{2\pi}{\oint_{\Gamma}{{\nabla_{\textbf k}} {\arg[E_m(\textbf k)-E_n(\textbf k)]}} \,d\textbf{k}},
 \end{equation}
 
where $\Gamma$ is a closed loop encircling the EP in the momentum space. The energy eigenvalue for a single complex band of non-Hermitian Hamiltonian in general can be written as $E(k)=\mathopen|E(k)\mathclose|e^{i\theta_L(k)}$, where $\theta_L=\tan^{-1}({\mathrm{Im}E/\mathrm{Re}E})$. The complex energy bands evolve in a periodic cycle $\theta_L(k)\rightarrow \theta_L(k)+2v\pi$ ($v$ being an integer). Since an EP is a defective point in the spectrum, encircling an EP leads to a quantized vorticity. For our photon dressed non-Hermitian nodal line semimetal case, we illustrate the topological transitions by calculating the evolution of the two bands encircling an EP at different values of light intensity. In Fig.~\ref{Vorticity_2} we present the vorticity of one of the EPs for zero light intensity and clearly see the swapping of two bands, indicating band degeneracy as a signature of nontrivial topological phase. Gradually increasing the light intensity leads to the deformation of evolution of the two complex bands. Eventually at a critical value of light intensity (see Eq.~\ref{critical_A-value}), when the four EP locations become imaginary we see the two energy bands get separated and topological properties of the system are lost, as there is no swapping between energy bands in the complex energy plane. Overall, our analysis shows that one can control the stability of EPs by tuning the light amplitude and eventually also annihilate them.Furthermore, in section VI, we will also show that light can engineer topological phase transitions from trivial to non-Hermitian topological phases creating exceptional points in the system.

\begin{figure*}
\includegraphics[scale=0.22]{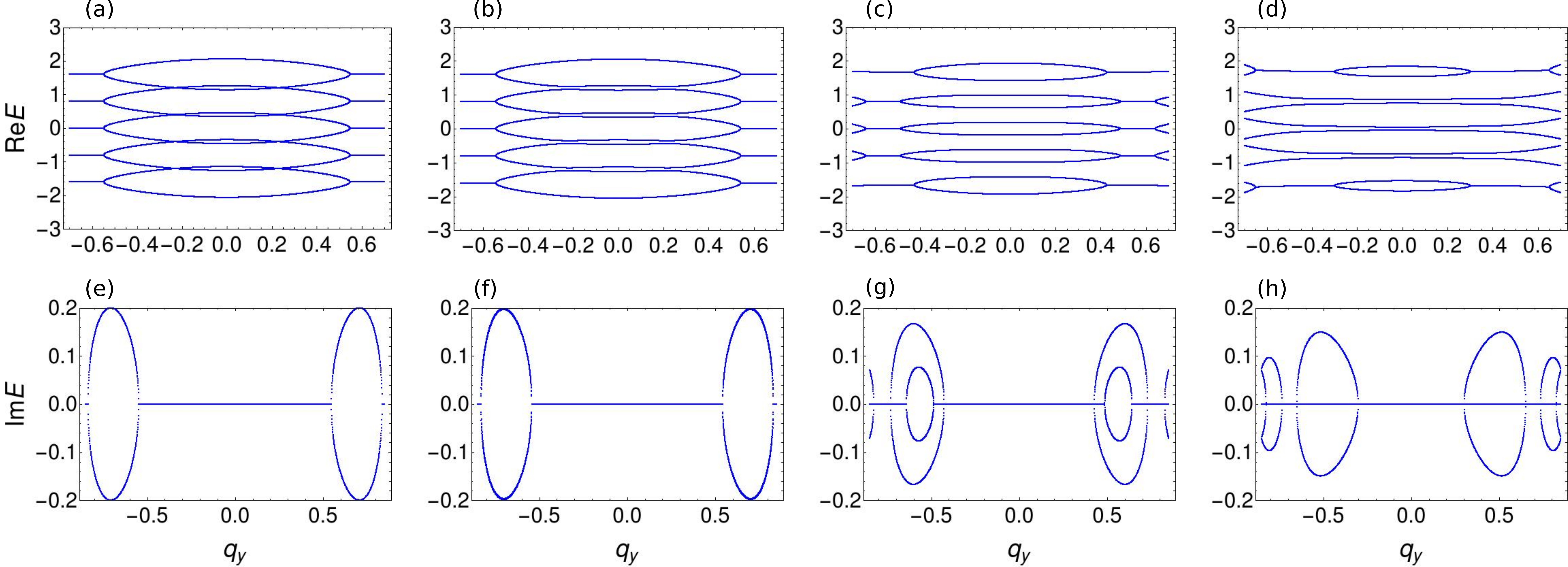}
  \caption{\textbf{Numerical analysis of Floquet eigenspectra as a function of light intensity.} The real and the imaginary parts of the energy dispersion (a) and (e) in the absence of light $(a_y = a_z = 0)$, (b) and (f) with moderate laser intensity $(a_y = a_z = 0.08$ and (c) and (g) at $(a_y = a_z = 0.35)$. The exceptional ring squeezes and the exceptional points come closer with increasing light intensity. At this critical light intensity $(a_y = a_z = 0.45)$, (d) the real and (h) imaginary part of the spectrum are shown. We observe the change in band topology as a function of light intensity, i.e., the shape of the rings change. Beyond the critical light intensity the exceptional points coalesce with each other resulting in the destruction of exceptional points in the spectrum. We choose the following values for the other parameters: $B = v_z = 1.0$, $e = 1$, $\Omega = 0.8$, $\eta = 1.0$, $m = 0.5$ and $\gamma = 0.2$. } \label{numerical_floquet}
\end{figure*}

\section{Numerical Analysis of Floquet equations without high frequency expansion}

In this section we study the effect of AC fields on the non-Hermitian topological nodal line semimetal and investigate the effect of light on the stability of exceptional points solving the full Floquet eigenvalue equations. The full time-dependent Hamiltonian in the presence of off-resonant light, which generates a vector potential $\textbf{A} (t)=a_y\eta\cos{\omega t}\hat{y}+a_z\cos{(\omega t +\phi)}\hat{z},$ becomes

\begin{equation}
\begin{split}
H(t)=[m-B(q_x^2+(q_y+a_y\eta\cos{\omega t})^2\\
+ (q_z+a_z\cos{(\omega t+\phi))^2}]\sigma_x \\
+[v_z(q_z+a_z\cos{(\omega t + \phi)})+i\gamma]\sigma_z.
\end{split}
\end{equation}

With Fourier transformed Floquet states $|\phi(t)\rangle=\sum_{m}e^{-i m \Omega t}|u_m^{\alpha}\rangle$ the Floquet eigenvalue equation reads\cite{oka2009photovoltaic}

\begin{equation}
\sum_n H^{mn} |u_m^{\alpha}\rangle=(\varepsilon_\alpha+m \Omega)|u_m^{\alpha}\rangle,
\end{equation}

where the Floquet state is labeled by $\alpha = (i, m)$ with $i = 1,2$ representing the upper and lower branches of the two band Hamiltonian, and $m$ is the Floquet index.
The matrix elements of the Floquet Hamiltonian are given by $ H^{mn} =\dfrac{1}{T} \int_0^{T} dt H(t) e^{i(m-n)\Omega t}$. In principle, following this approach one needs to diagonalize an infinite dimensional matrix. However, numerically one resorts to choosing a few values of $m$~\cite{oka2009photovoltaic,usaj2014irradiated}. Here we chose $m=-2,-1,0,1,2$.
We present our numerically computed eigenspectra of the full Floquet Hamiltonian in Fig.~\ref{numerical_floquet}. We plot the real and imaginary parts of the eigenenergy along $q_y$. At zero energy, we see the cross section of exceptional rings in the spectrum. With increasing light intensity the topology of the ring changes. The motion of exceptional points is evident as the ring squeezes with increasing light intensity and eventually they get annihilated beyond a critical intensity resulting in a gap in the spectrum. This numerical analysis further confirms our inferences from the high frequency expansion presented in section II. \\

\section{Effect of light under open boundary conditions}

\begin{figure*}
\includegraphics[scale=0.35]{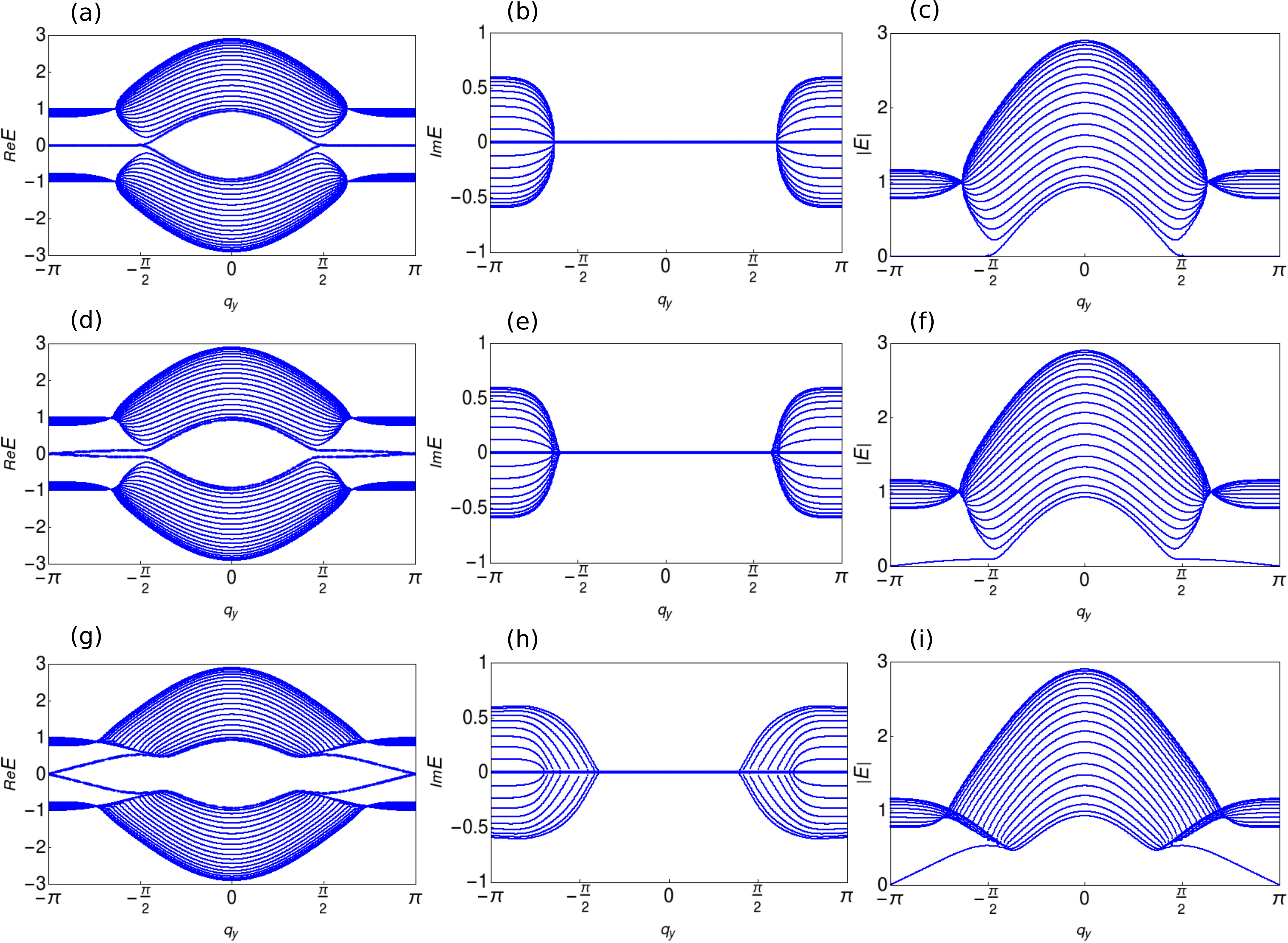}
  \caption{\textbf{Analysis of the nodal line semimetal lattice model under open boundary conditions.} (a) The real, (b) imaginary and (c) absolute part of the energy dispersion in the absence of light ($a_y=a_z=0$). We find dynamically stable edge modes and higher order exceptional point in the spectrum. The real, imaginary and absolute parts of the energy dispersion in the presence of light (d-f) at moderate light intensity ($a_y=a_z=0.65$) and (g-i) at higher light intensity ($a_y=a_z=0.8$). We see that the boundary modes have moved off zero energy, rendering them unstable. The stability of higher order exceptional point is also destroyed. Here $m=3.0$, $B=0.5$, $v_z=1.0$,  $\gamma=0.6$, $e=1.0$, $\eta=1.0$, and $\omega=1.0$. keeping them unchanged unless otherwise specified.} \label{nodal_edge}
\end{figure*} 

\begin{figure*}
\includegraphics[scale=0.4]{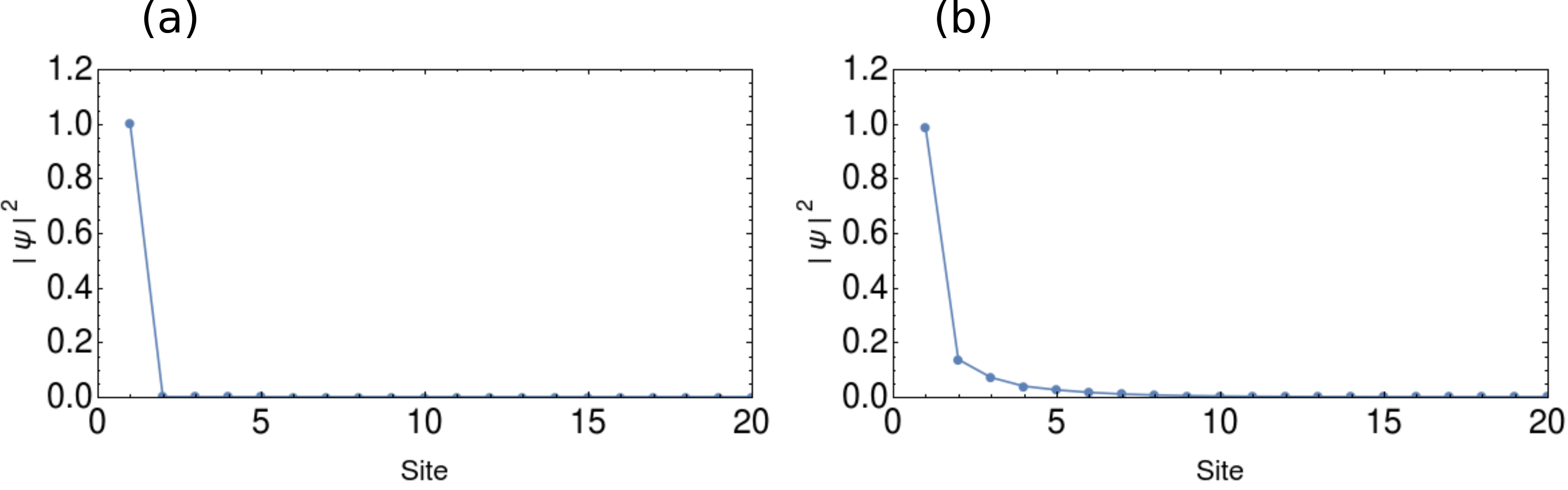}
  \caption{\textbf{Localization of the wave function under open boundary conditions.} The wave function corresponding to the boundary mode at $q_y=0.6\pi$, for (a) no illumination and (b) $a_y=a_z=0.6$. The wave function corresponding in the absence of light is highly localized at one surface. With increasing light intensity the surface localization decreases as the boundary states moved off from zero energy.} \label{nodal_edge_wavefunction}
\end{figure*}

\begin{figure*}
\includegraphics[scale=0.35]{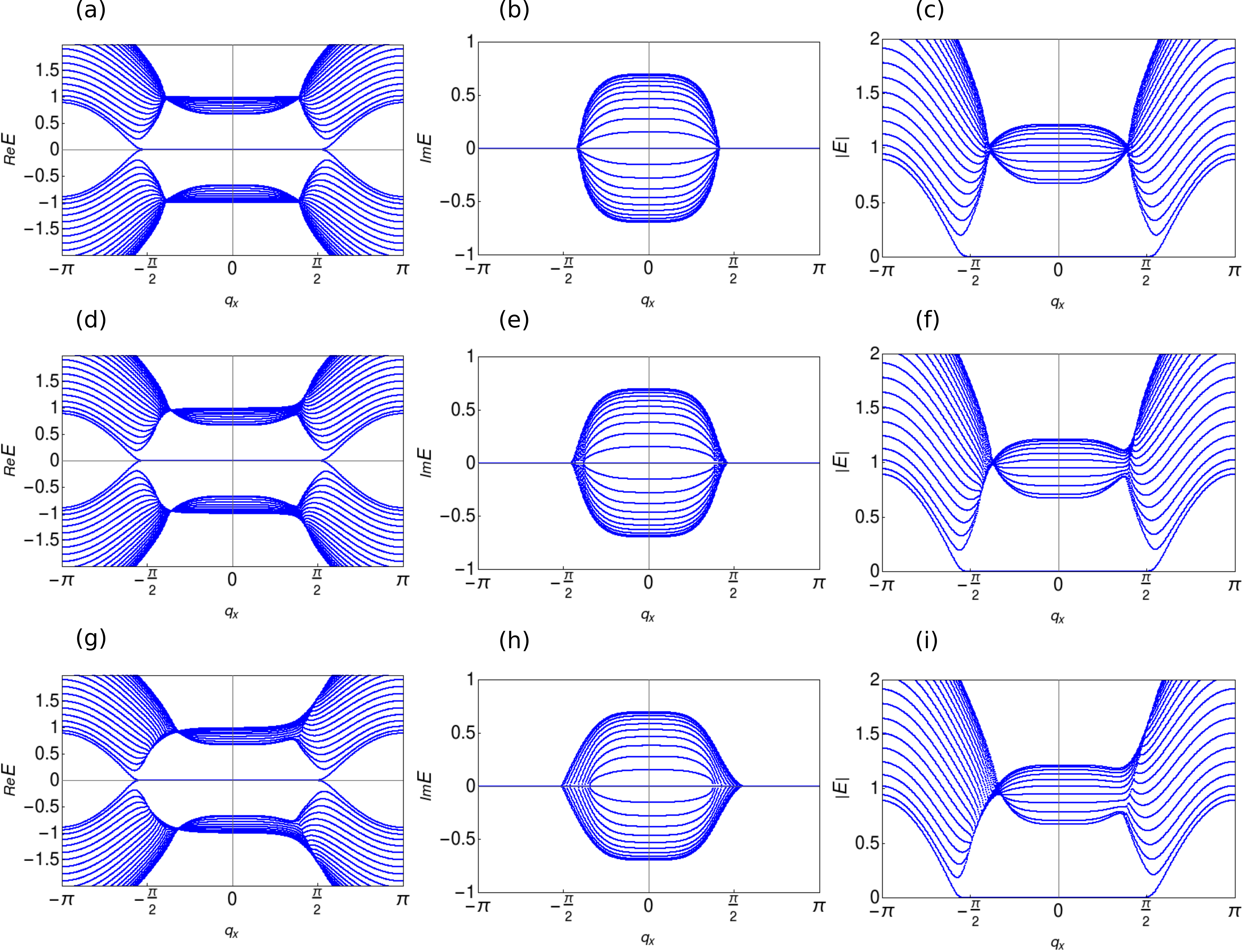}
  \caption{\textbf{Analysis of the semi-Dirac semimetal lattice model under open boundary conditions.} The real (left panel), imaginary (middle panel) and absolute (right panel) parts of the spectrum under (a-c) no illumination, (d-f) moderate light intensity ($a_x=a_y=0.2$), and (g-i) relatively strong light intensity ($a_y=a_z=0.4$). We find the higher order exceptional points are stable for zero light intensity. The stability of higher order exceptional points are destroyed for finite light intensity. We have chosen following parameters: $M=1.0$, $t_1=1.0$ and $v_f=1.0$.} \label{semidirac_lattice}
\end{figure*}

In this section, we study the effect of off-resonant light on zero-energy drumhead-like surface bands and higher order exceptional points obtained from the lattice model corresponding to nodal line semimetal Hamiltonian (Eq.~\ref{model_Hamiltonian}), when opened along the $z$ direction. From Eq.~\ref{model_Hamiltonian} following Ref.~\onlinecite{wang2019non}, the lattice model Hamiltonian can be obtained as

\begin{equation}\label{lattice_Hamlitonian}
H= [m-2B(3-\cos{q_x}-\cos{q_y}-\cos{q_z})]\sigma_x + (v_z \sin{q_z} +i \gamma)\sigma_z.
\end{equation}

We choose the vector potential to be $\textbf{{A}}(t)=a_y\eta \cos{\omega t}\hat{y}+a_z\sin{\omega t}\hat{z}$, which is the same as for the continuum model with $\phi=\pi/2$. Using the Floquet formalism (Eq.~\ref{effecive_floquet}) and considering the minimal coupling $\textbf{q}\rightarrow \textbf{q} + e\textbf{{A}}(t)$ we obtain light induced term as

\begin{equation}
    H_{\mathrm{floquet}}=-\dfrac{8 B v_z J_1(e \eta a_y) J_1(e  a_z)}{\omega} \sigma_{y} \sin{q_y} \cos{q_z},
\end{equation}

where $J_{n}(x)$ denotes the $n$-th order Bessel function. To get the surface states, we choose open boundary conditions in the $z$ direction to numerically calculate the spectrum as a function of $q_y$ for 40 sites. We present the real, imaginary and absolute part of the spectra for a fixed $q_x=0$ in Fig.~\ref{nodal_edge}. In the absence of light we find the higher order EP, where all eigenvalues coalesce. We get dynamically stable zero modes since absolute part of the flat bands equals zero~\cite{wang2019non}. Next, we turn on light to observe the effect of light on these stable zero modes as well as the EP. We find that the stability of zero energy modes is destroyed on introduction of light as they move away from zero energy. Furthermore, we find that the stability of higher order EP is also destroyed. So these findings suggest that the results from the lattice model under open boundary conditions are in good agreement with the continuum model. We have also plotted the wave function associated with the boundary state in Fig.~\ref{nodal_edge_wavefunction}. We find that the wave function is well localized at the boundary in the absence of light. With increasing light intensity the wave function decays slower into the bulk as the boundary modes move away from zero energy.

\section{Semi-Dirac semimetal in two dimensions} 

As a second example of our proposal, we consider a non-Hermitian semi-Dirac semimetal in two dimensions described by the Hamiltonian~\cite{banerjee2020non}

\begin{equation}
H(\textbf q)=\left(\frac{q_x^2}{2m}-\delta_0\right)\sigma_x+v_fq_y\sigma_z+i\gamma\sigma_z.
\end{equation}

\noindent Here also $\gamma$ is the gain and loss coefficient between the two orbitals. The Pauli matrices act on orbital subspace. Here $\mathbf{q}$ is the crystal momentum, $m$ is the quasi-particle mass along the quadratically dispersing direction, $v_f$ is the Dirac velocity, and $\delta_0$ is the gap parameter. We now introduce circularly polarized off-resonant light, with a vector potential $\textbf{{A}}(t)=a(\eta \sin{\omega t} \hat{x}+ \cos{\omega t} \hat{y})$. Using Floquet formalism discussed previously, we obtain the effective Hamiltonian in the presence of light as

\begin{equation}
H_{\mathrm{eff}}=\left(\frac{q_x^2}{2m}-\delta_0\right)\sigma_x+(v_f q_y +i \gamma)\sigma_z - \dfrac{\eta e^2 a^2 v_f q_x}{m \omega}\sigma_y.
\end{equation}

We have evaluated the effect of light on the EPs in this two-dimensional system (see appendix B for detailed calculations). The modified positions of the EPs, due to light, along the nodal line $q_y=0$ become
 
\begin{equation}
q_{\mathrm{EP}}=\pm\sqrt{m} \sqrt{(2 \delta_0 -2m\xi^2)\pm\sqrt{(2\delta_0 - 2m\xi^2)^2-4(\delta_{0}^2-\gamma^2)}},
\end{equation}

where $\xi=\dfrac{\eta e^2 a^2 v_f}{m \omega}$. We present the locations of EPs for different values of $a$ in upper panels of Fig.~\ref{Vorticity_3}. We discover that one can indeed tune the positions of EPs by changing the laser intensity. We find a critical laser intensity, at which the four EPs annihilate each other pairwise, rendering their positions imaginary, as

\begin{equation}
\begin{split}
\xi_c & = \sqrt{\dfrac{2\delta_0\pm 2\sqrt{(\delta_{0}^2-\gamma^2)}}{2m}}, \\
a_c & =\left(\dfrac{m^2\omega^2}{\eta^2 e^4 v_f^{2}}\right)^{1/4} \left({\dfrac{2\delta_{0}\pm 2\sqrt{(\delta_{0}^2-\gamma^2)}}{2m}}\right)^{1/4}.
\end{split}
\end{equation}

Next, we characterize our system under illumination using the vorticity. We illustrate the topological changes of the system in Fig.~\ref{Vorticity_3} for increasing values of light intensity. We find a change in evolution of two bands as $a$ is increased. Two pairs of the EPs approach each other and mutually annihilate. So, we have found that the laser intensity can be used in topological tuning and the accompanying topological phase transitions in two-dimensional semi-Dirac semimetals. 

We next study the effect of light on exceptional points under open boundary conditions. The lattice model corresponding to the low energy semi-Dirac dispersion reads

\begin{equation}
    H(q)=(m_x-t_2\cos{q_y})\sigma_x+(v_f\sin{q_y+i\gamma})\sigma_y,
\end{equation}

bluewhere $m_x=M-t_1\cos{q_x}$, $M=(1/m-\delta_0)$, $t_1=1/m$ and we further choose $t_2=v_f$. We choose the vector potential to be $\textbf{{A}}(t)=a_x\eta \sin{\omega t}\hat{x}+a_y\cos{\omega t}\hat{y}$, the same as for the continuum model. Using the Floquet formalism (Eq.~\ref{effecive_floquet}) and considering the minimal coupling $\textbf{q}\rightarrow \textbf{q} + e\textbf{{A}}(t)$ we obtain light induced term as

\begin{equation}
    H_{\mathrm{floquet}}=-\dfrac{4 v_f t_1 J_1(e \eta a_x) J_1(e  a_y)}{\omega} \sigma_{y} \sin{q_x} \cos{q_y},
\end{equation}

where $J_{n}(x)$ denotes the $n$-th order Bessel function. The light induced term is proportional to $\sin{q_x}$. Here we choose open boundary conditions in the $y$ direction to numerically calculate the spectrum as a function of $q_x$. We present the real, imaginary and absolute parts of the spectra in Fig.~\ref{semidirac_lattice}. In the absence of light we find a higher order EP, where all the eigenvalues coalesce. With increasing light intensity we observe that the stability of the higher order EP tends to be destroyed, which is in good agreement with our low energy model results.

\begin{figure*}
\includegraphics[scale=0.3]{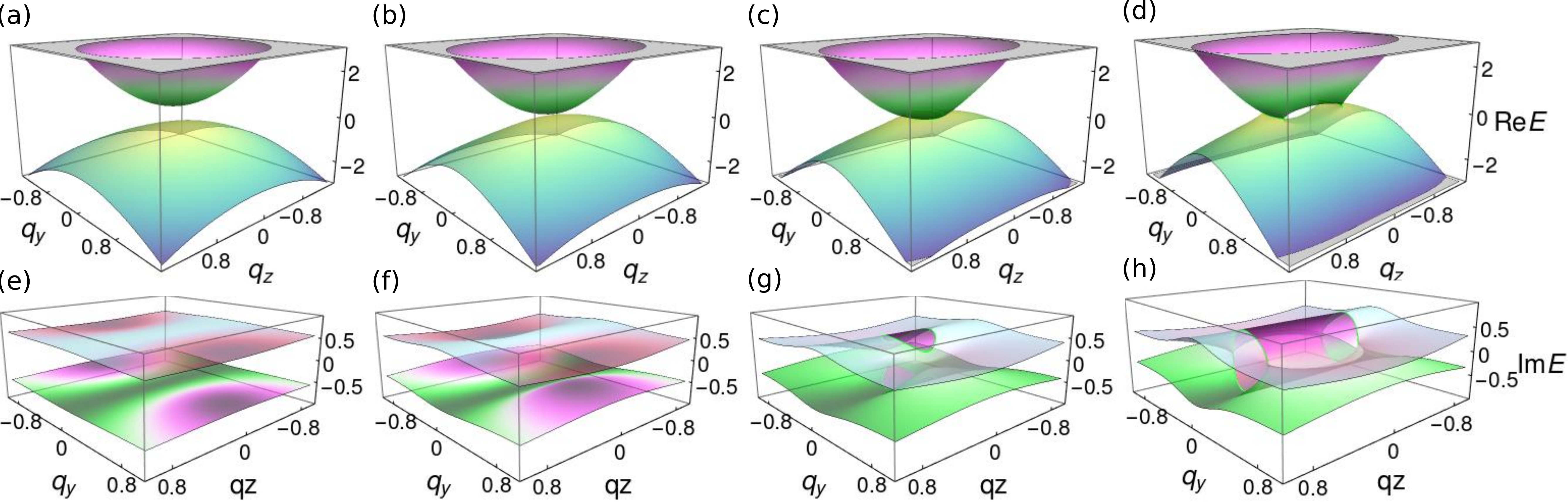}
  \caption{\textbf{Creating exceptional points in three-dimensional non-Hermitian Dirac semimetal using Floquet engineering.} The real and the imaginary part of the energy dispersion (a) and (e) in the absence of light ($a=0$), (b) and (f) with moderate laser intensity ($a=0.6$), and (c) and (g) at critical light intensity ($a=0.75$). At the critical value the two bands touch each other. Beyond this critical light intensity, (d) the real and (h) imaginary part of the spectrum are shown. We obtain four exceptional points in the spectrum. Note that in the imaginary spectrum, two bands get exchanged. We choose the following values for the other parameters: $c_0=0$, $c_1=c_2=0.5$, $m_0=0.5$, $m_1=m_2=-1.0$, $\gamma=0.6$, $q_x=0.0$, $\eta=1.0$, $A=1.0$ and $\omega=1.0$, keeping them unchanged unless otherwise specified.} \label{band_diagram_dirac}
\end{figure*}

\begin{figure*}
\includegraphics[scale=0.35]{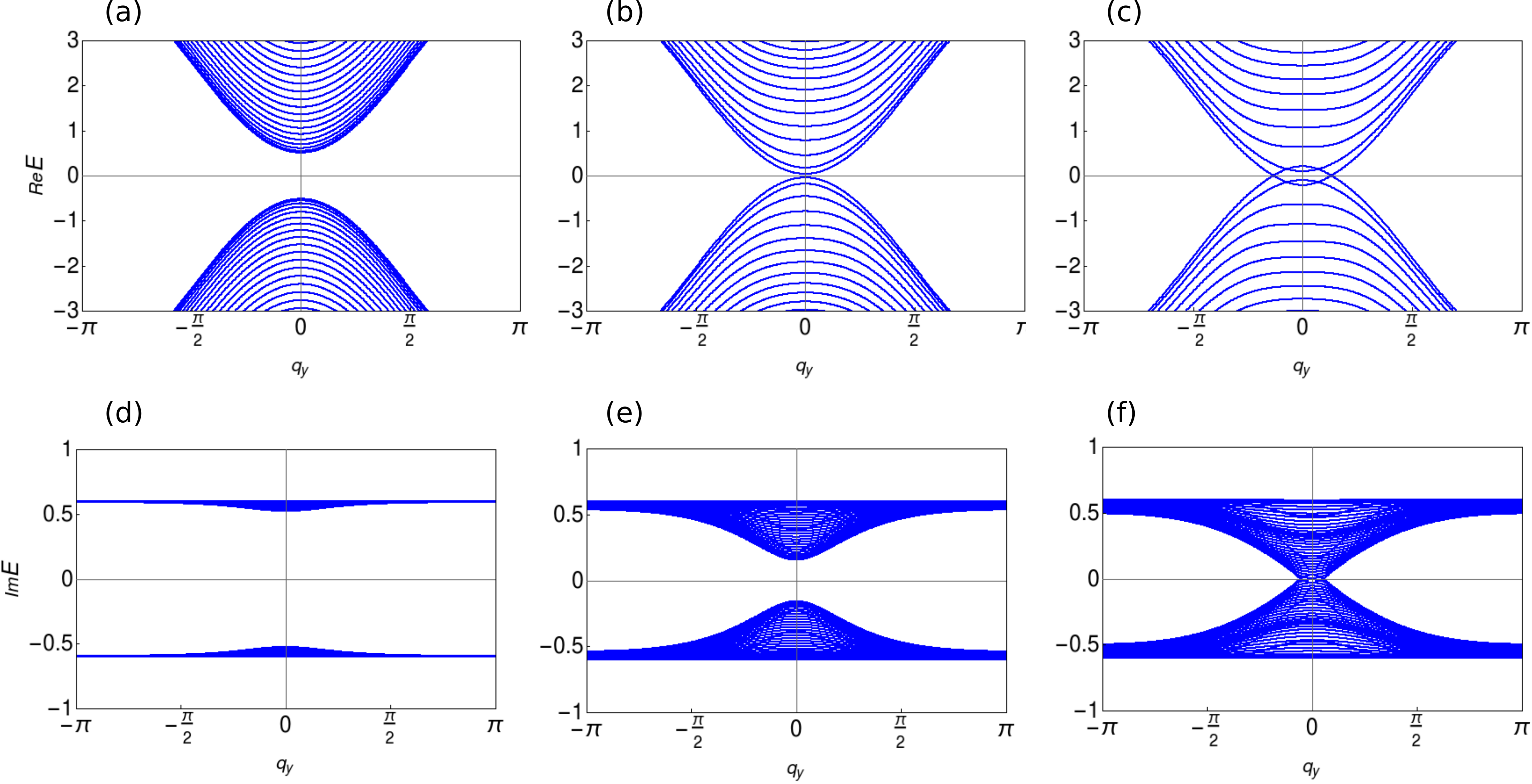}
  \caption{\textbf{Analysis of the Dirac semimetal lattice model under open boundary conditions.} (a) and (d) show the gapped real and imaginary spectra, respectively, in the absence of light. With increasing light intensity ($a=0.65$), the gap decreases as shown in (b) and (e). (c) and (f) Beyond the critical light intensity ($a=0.85$), we obtain exceptional points in the spectra enabling topological phase transitions.} \label{Diarc_edge_states}
\end{figure*}

\section{Creating Exceptional Points In Dirac Semimetals Using Floquet Engineering}

So far we presented our results showing the control over the position of exceptional points. In this section, we elucidate a proposal for \emph{creating} exceptional points using Floquet engineering. We investigate the effect of off-resonant light on three-dimensional non-Hermitian Dirac semimetals. A typical three-dimensional Hermitian Dirac semimetal can be described by the continuum model Hamiltonian~\cite{wang2012dirac,wang2013three,narayan2015floquet}

\begin{equation}
H=\epsilon_0(\textbf{k}) \tau_0+M(\textbf{k})\tau_z+
A(q_x\tau_x-q_y\tau_y),
\end{equation}

where $\epsilon_0(\textbf{k})=c_0+c_1q_z^2+c_2(q_x^2+q_y^2)$, $M(\textbf{k})=m_0-m_1q_z^2-m_2(q_x^2+q_y^2)$ and $\tau_i$ $(i=x,y,z)$ are Pauli matrices acting in the two-orbital subspace, $\tau_0$ is the identity matrix. $M(\textbf{k})$ is the band inversion parameter. The energy can be obtained as  $E_{\pm}=\epsilon_0(\textbf{k})\pm \sqrt{M(\textbf{k})^2+A^2(q_x^2+q_y^2)}$. We get Dirac crossings at $q_c=(0,0,q^{c}_z=\pm\sqrt{m_0/m_1)}$, when the signs of $m_0$ and $m_1$ are the same (else one obtains a fully gapped spectrum). We next introduce a non-Hermitian term $i\gamma\tau_z$ $(\gamma >0)$ associated with particle gain and loss for the two orbitals. The Hamiltonian now reads

\begin{equation}\label{dirac_floqet}
H=\epsilon_0(\textbf{k}) \tau_0+M(\textbf{k})\tau_z+
A(q_x\tau_x-q_y\tau_y)+i\gamma\tau_z.
\end{equation}

The energy becomes complex in general. Now, we study the effect of off-resonant circularly polarized light which generates a vector potential $a(t) = a(\eta\sin{\omega t}\hat{x}+\cos{\omega t}\hat{y})$ on this non-Hermitian system. Following the Floquet approach outlined previously, we find that the commutator takes the form $[H_{-1},H_{+1}]=-2\eta e^2a^2 A m_2(q_x\tau_x-q_y\tau_y)-\eta e^2a^2A^2\tau_z$. Incorporating the light induced term, the effective Hamiltonian becomes

\begin{equation}
\begin{split}
H=\epsilon_0(\textbf{k}) \tau_0+M(\textbf{k})\tau_z+ A(q_x\tau_x-q_y\tau_y)+i\gamma\tau_z+\\
\eta (\dfrac{-2 e^2 a^2 A m_2 q_x}{\omega} \tau_x + \dfrac{2 e^2 a^2 A m_2 q_y}{\omega}\tau_y+ \dfrac{- e^2 a^2 A^2}{\omega} \tau_z).
\end{split}
\end{equation}

The light induced terms proportional to $\tau_x$ and $\tau_y$ are linear in momentum and modify the velocities in the $x,y$ directions, whereas the light induced term proportional to $\tau_z$  effects the band inversion strength. We present the band diagram in Fig.~\ref{band_diagram_dirac} starting from topologically trivial insulator ($m_0>0$, $m_1<0$, $\gamma>0$ and $a=0$). We notice that, with increasing light intensity, the band gap decreases, reaching a critical zero gap. So, light can change the Fermi surface effectively and beyond some critical light intensity, we discover four exceptional points in the spectra. We note the exchange of two bands forming Riemann sheet geometry in the imaginary spectra. We can also tune the positions of exceptional points changing the light intensity. Thus, depending on the polarization of light, one can create and tune positions of exceptional points in this non-Hermitian system.  \\

Next, in order to investigate the effect of illumination under open boundary conditions, we switch to the lattice model corresponding to Eq.~\ref{dirac_floqet} by making the replacements $k\rightarrow \sin k$ and $k^2\rightarrow 2(1-\cos k)$. The light induced Floquet term is found to be

\begin{equation}
H_{floquet}=\eta  \dfrac{2 e^2 a^2 A  m_2 \sin{q_y}}{\omega}\tau_y- \dfrac{ e^2 a^2 A^2}{\omega} \tau_z.
\end{equation}

We present the numerically obtained spectra as a function of $q_y$ in Fig.~\ref{Diarc_edge_states}. In the absence of illumination, we have a gapped spectrum with open boundary conditions along $z$. With increasing light intensity the gap in both the real and imaginary part of the energy decreases. Beyond a critical light intensity the gaps close and we obtain EPs. Therefore, the lattice model confirms our results from the low energy model and Dirac semimetals can be a useful platform to engineer EPs under illumination.

\section{Possible experimental realization} 

Over the last few years, there have been several exciting experiments suggested, as well as realized, to explore rich topological phases by tuning the gain and loss terms in a controllable manner, mainly in dissipative waveguide systems and cold atomic gases~\cite{xu2017weyl,midya2018non,alvarez2018topological}. Furthermore many ingenious experiments have been devoted to study the topological signatures of exceptional points~\cite{dembowski2001experimental,ding2016emergence}. 

Combining the ideas of circular modulation of waveguides~\cite{rechtsman2013photonic} and insertion of breaks to achieve gain and loss~\cite{cerjan2019experimental}, we expect that our proposal of tuning EPs in two-dimensional models and two-dimensional slices of three-dimensional models should be within experimental reach. In addition, our proposal may also be realized in ultra-cold atomic setups, where shaking of the optical lattice can provide a drive~\cite{jotzu2014experimental}, while atomic population control can yield the non-Hermiticity~\cite{li2019topology}. Additionally, topological laser systems can be potential hosts to study the light induced topological properties of EPs~\cite{bandres2018topological,st2017lasing}. 

Next, we discuss the possible experimental realization of our formalism and estimate various parameters related to the system and their experimental feasibility in existing technologies. Following Eq.~\ref{effective-hamiltonian}, we see $\zeta$ has dimensions of velocity and it scales with bulk Fermi velocity of the system. Consider a photonic lattice consisting of helical waveguides with a pitch ($Z=1$ cm). For a given helix radius of the waveguide ($R=1$ $\mu$m) and refractive index $n_0=1.45$, at a typically used wavelength ($\lambda=650$ nm), one can estimate $ea \approx 87.5$ cm$^{-1}$. $B$ has the dimensions of inverse energy, so for a typical choice $B^{-1}= 0.8$ eV  (or $6450$ cm$^{-1}$), we find $\zeta \approx 0.37v_z$. This fraction of bulk Fermi velocity is experimentally feasible in existing wave-guide technologies~\cite{rechtsman2013photonic}. One can tune the light intensity by choosing different helix radius, as has been experimentally achieved~\cite{rechtsman2013photonic,cerjan2019experimental}. On the other hand, if one considers slightly different values $R=4$ $\mu$m and $B^{-1}=2$ eV, then $\zeta$ becomes $0.6v_z$. Therefore, the choice of the helix radius allows one to readily modulate the effective Floquet term in this setup.

Having achieved the Floquet part of our proposal, we move on to the gain and loss terms. We can introduce gain and loss in the system by incorporating breaks between waveguides, as has been previously demonstrated experimentally~\cite{cerjan2019experimental}. If we consider the typical break-length to be $40$ $\mu$m, a typical refractive index shift $\Delta n=2.6 \times 10^{-3}$ and a helix radius ($R=4$ $\mu$m) one can estimate the strength of gain and loss coefficient to be 0.2. Furthermore, one can tune the strength of gain-loss by increasing the break-length between waveguides~\cite{cerjan2019experimental}.

An alternative approach to the possible experimental realization of our proposal is by illuminating a topological insulator-ferromagnet junction -- a system which has been recently theoretically identified as a platform for realizing non-Hermitian gapless phases by Bergholtz and Budich~\cite{bergholtz2019non}. This platform exhibits exceptional points and shining light on the junction would be a feasible solid-state approach to realize our proposal for manipulating exceptional points. Overall, given the rapid advancements in experiments in the field of non-Hermitian topological physics, we are optimistic about the prospects for the experimental study of our proposals.

\section{Summary and conclusions} 

In summary, we have proposed a new platform for controlling exceptional points in non-Hermitian topological systems using light. Using three different non-Hermitian examples -- nodal line semimetals, semi-Dirac semimetals and Dirac semimetals -- we have shown that light can create as well as tune the positions and stability of exceptional points. Furthermore, one can generalize our formalism beyond these examples to other non-Hermitian systems which can be tuned using light. We hope that these findings will motivate exploration of using light to engineer exceptional points in non-Hermitian systems.

\section{Acknowledgments} We thank Dr. Vivek Tiwari for useful comments on the manuscript. A.B. is supported by a fellowship from the Indian Institute of Science. A.N. acknowledges support from the start-up grant (SG/MHRD-19-0001) of the Indian Institute of Science.

\section{APPENDIX A: Light-induced topological properties with different types of non-Hermitian terms}

\begin{figure*}
 \includegraphics[scale=0.50]{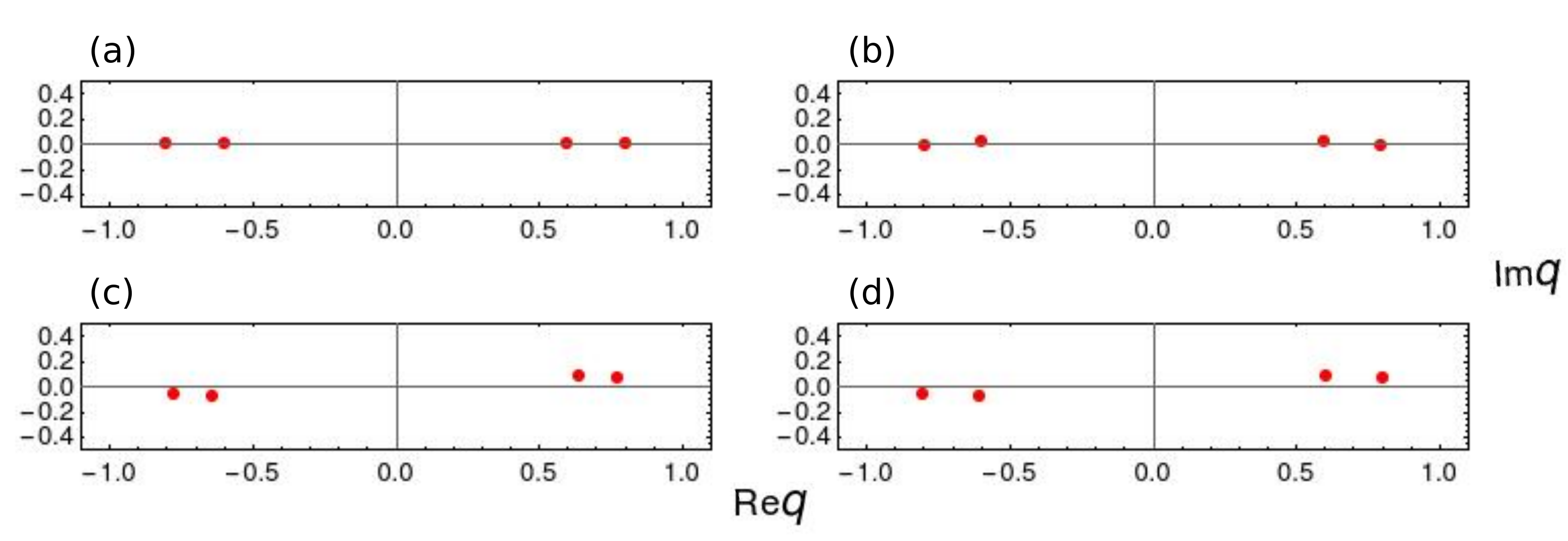}
\caption{\textbf{Locations of exceptional points of photon dressed non-Hermitian nodal semimetals considering different types of non-Hermitian perturbation in the complex plane.} The upper panels show the location of exceptional points in the (a) absence $(a_y=a_z=0) $ and (b) presence of $(a_y=a_z=0.15)$ light with only $\gamma_2=\gamma_3 \ne 0 $. In this case light instantaneously destroys the exceptional points as we see in (b). The locations of exceptional points in the absence of light are shown in (c) for $\gamma_1=\gamma_3 \ne 0 $ and (d) $\gamma_1=\gamma_2=\gamma_3 \ne 0 $ respectively. Both these cases lack exceptional points. We choose the following values for the other parameters: $B=v_f=1.0$, $e=\omega=1.0$, $m=0.5$, $\gamma_1=\gamma_2=0.1$ and $\gamma_3=0.1$.}\label{ep-plot-non-hermitian}
 \end{figure*}

In this section, we present light-induced spectra of nodal line semimetal considering both complex hopping and complex onsite potentials. Although our main motivation was to study the light-induced topological phase transition using non-Hermitian gain and loss potential, i.e., the non-Hermitian term proportional to $\sigma_z$, as it is experimentally achievable and well reported in the literature~\cite{cerjan2019experimental,midya2018non}. We focused on this form of the non-Hermitian term in the main text. In principle, one can include three kinds of non-Hermitian terms in the form of $H_{NH}=i\mathbf{\gamma_j}\cdot\mathbf{\sigma_j}$ where $j=1,2,3$ ($\sigma_j$ are the Pauli matrices). In our model we find that inclusion of a non-Hermitian term proportional to $\sigma_1$ case has no topological properties and lacks exceptional points. In the absence of light, with combined $\sigma_2$ and $\sigma_3$ terms, the system shows exceptional points. We find that light instantaneously destroys the exceptional points opening a gap in the spectrum. We present the locations of exceptional points in Fig.~\ref{ep-plot-non-hermitian}. The imaginary coordinates of exceptional points in the momentum space indicate their non-existence in the presence of light. We have also considered other non-Hermitian perturbations such as a combination of terms proportional to both $\sigma_1-\sigma_3$ and $\sigma_1-\sigma_2-\sigma_3$. We find that these are not topological and lack of exceptional points in (c) and (d). The stability of exceptional points is not restored even in the presence of light. Nevertheless, we emphasize that it is experimentally possible to control which form of non-Hermitian term one introduces in the system~\cite{cerjan2019experimental,midya2018non}.
 
Our observations can be explained as follows. The Hamiltonian in the present case can be written as $H(k)=\mathbf{d}(k)\cdot \boldsymbol{\sigma}$, where $\boldsymbol{d}=\mathbf{d}_R +i \mathbf{d}_I$ with $\mathbf{d}_R,\mathbf{d}_I \in \mathfrak{R}^3$ and $\boldsymbol{\sigma}$ the vector of standard Pauli matrices. To get the locations of exceptional points the following two equations need to be satisfied simultaneously
\begin{eqnarray}
    d_R^{2} - d_I^{2} =0, \\
    \mathbf{d}_R \cdot \mathbf{d}_I = 0.
\end{eqnarray}

The latter one signifies that nonzero light intensity can destroy the stability of exceptional points with combined terms proportional to $\sigma_2$ and $\sigma_3$ terms. We do not get exceptional points for other cases following these two conditions. We can also draw similar conclusions for the non-Hermitian semi-Dirac case. 

\begin{figure*}
\includegraphics[scale=0.35]{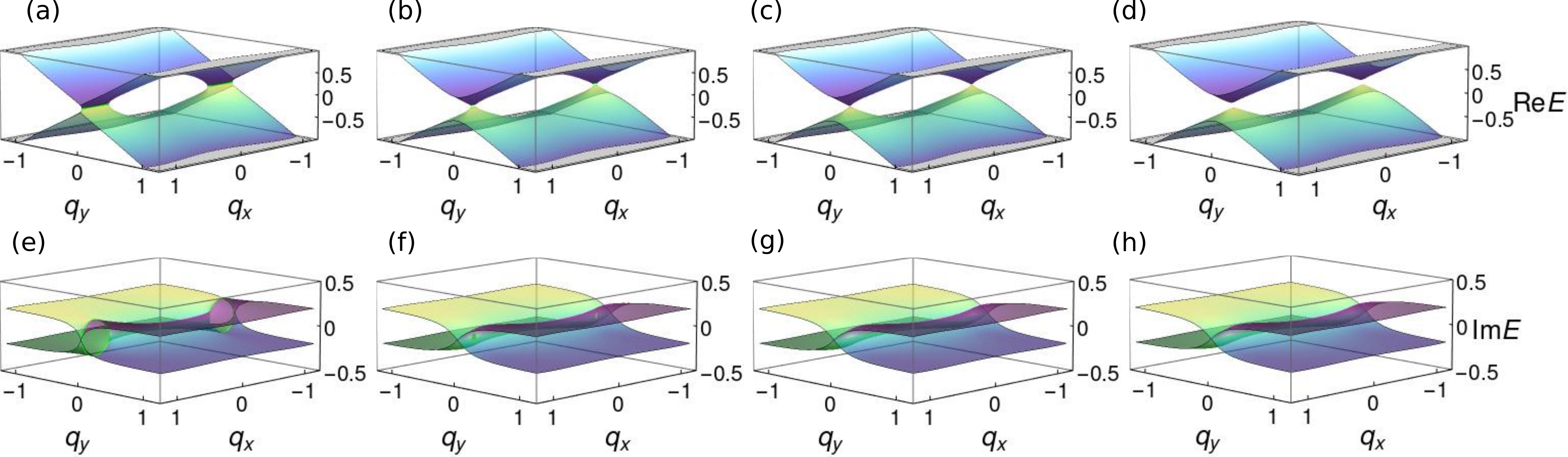}
  \caption{\textbf{Band diagrams of photon-dressed non-Hermitian semi-Dirac semimetals.} The real and the imaginary part of the energy dispersion (a) and (e) in the absence of light ($a=0$), (b) and (f) with moderate laser intensity ($a=0.45$), and (c) and (g) at critical light intensity ($a=0.452$). At the critical value the nodal line symmetry of the spectrum is lost along the $q_y=0$ line. Beyond this critical light intensity, (d) the real and (h) imaginary part of the spectrum are shown. The real part of the spectrum become gapped. We choose the following values for the other parameters: $B=v_f=1.0$, $e=\omega=1.0$, $m=1$, $\eta=1$, $\delta=0.5$ and $\gamma=0.2$.} \label{band_diagram_semidirac}
\end{figure*}

\section{APPENDIX B: Floquet engineering of non-Hermitian semi-Dirac semimetals}

In this section, we present further calculation details of Floquet analysis of a semi-Dirac semimetal. The model Hamiltonian describing low-energy electronic bands of a two-dimensional non-Hermitian semi-Dirac~\cite{banerjee2020non} semi-metal incorporating a gain and loss term between the two orbitals is

\begin{equation} \label{Non-Hermitian_Hamiltonian}
H(\textbf q)=\left(\frac{q_x^2}{2m}-\delta_0\right)\sigma_x+v_fq_y\sigma_z+i\gamma\sigma_z,
\end{equation}

\noindent where $\boldsymbol{\sigma}= (\sigma_x, \sigma_y, \sigma_z)$ are the Pauli matrices in the pseudospin space and $\gamma$ can be thought of as the gain and loss coefficient between the two orbitals.

\par We get the energy eigenvalues as

\begin {equation} \label{energy_eigenvalues}
E=\pm\sqrt{\left(\frac{q_x^2}{2m}-\delta_0\right)^2+v_f^2q_y^2-\gamma^2+2iv_fq_y\gamma}.
\end{equation}

As a consequence of non-Hermitian band degeneracy, we find a nodal line along $q_y=0$ consisting of four EPs with symmetrically displaced locations given by
 
\begin{equation} \label{location_EP}
q_{x}^{{EP}}=\pm \sqrt{2m(\delta_0\pm\gamma)}.
\end{equation}

We now study the effect of circularly polarized off-resonant light of frequency $\omega$, which generates a vector potential $\textbf{{A}}(t)=a(\eta \sin{\omega t} \hat{x}+ \cos{\omega t} \hat{y})$, on this non-Hermitian semi-Dirac semimetal. Here $\eta=\pm 1$ for right and left circularly polarized beams, respectively. Using Floquet formalism discussed in the main text, we compute the light induced term for the Hamiltonian in Eq.~\ref{Non-Hermitian_Hamiltonian} as

\begin{equation}
[H_{-1},H_{+1}]=- \dfrac{\eta e^2 a^2 v_f q_x}{m} \begin{pmatrix}
0 & -i \\
i & 0
\end{pmatrix},
\end{equation}

\noindent such that the effective Hamiltonian in the presence of light becomes

\begin{equation}
H_{\mathrm{eff}}=\left(\frac{q_x^2}{2m}-\delta_0\right)\sigma_x+(v_f q_y +i \gamma)\sigma_z - \dfrac{\eta e^2 a^2 v_f q_x}{m \omega}\sigma_y.
\end{equation}

Note that similar to the nodal line semimetal, the perturbing term in the effective Hamiltonian is proportional to momentum, i.e., it is linear in $q_x$. We see in the band diagram presented in Fig.~\ref{band_diagram_semidirac}(a) that without the light induced term there are four EPs along the nodal line $q_y=0$. With increasing light intensity Fig.~\ref{band_diagram_semidirac}(b), (c) and (d), the positions of the EPs shift and eventually they annihilate each other resulting in a gap in the spectrum.

The dispersion relation for the effective Hamiltonian becomes

\begin {equation} \label{energy_eigenvalues2}
E=\pm\sqrt{\left(\frac{q_x^2}{2m}-\delta_0\right)^2+v_f^2q_y^2-\gamma^2+2iv_fq_y\gamma+\left(\dfrac{\eta e^2 a^2 v_f q_x}{m \omega}\right)^2}.
\end{equation}

The positions of the EPs along the nodal line $q_y=0$ change to

\begin{equation} \label{location_EP_floquet_1}
q_{\mathrm{EP}}=\pm\sqrt{m} \sqrt{(2 \delta_0 -2m\xi^2)\pm\sqrt{(2\delta_0 - 2m\xi^2)^2-4(\delta_{0}^2-\gamma^2)}},
\end{equation}

where $\xi=\dfrac{\eta e^2 a^2 v_f}{m \omega}$. In the main text, we have presented the locations of EPs for different values of $a$ and we have shown that we can tune the positions of EPs by changing the laser intensity. We find the critical laser intensity, at which the four EPs annihilate each other pairwise, rendering their positions imaginary as

\begin{equation}\label{critical_light_intensity}
\begin{split}
\xi_c & = \sqrt{\dfrac{2\delta_0\pm 2\sqrt{(\delta_{0}^2-\gamma^2)}}{2m}}, \\
a_c & =\left(\dfrac{m^2\omega^2}{\eta^2 e^4 v_f^{2}}\right)^{1/4} \left({\dfrac{2\delta_{0}\pm 2\sqrt{(\delta_{0}^2-\gamma^2)}}{2m}}\right)^{1/4}.
\end{split}
\end{equation}

So, the laser intensity can be used in topological tuning accompanying topological phase transitions. This result is confirmed by our vorticity analysis presented in the main text.


\begin{thebibliography}{66}%
\makeatletter
\providecommand \@ifxundefined [1]{%
 \@ifx{#1\undefined}
}%
\providecommand \@ifnum [1]{%
 \ifnum #1\expandafter \@firstoftwo
 \else \expandafter \@secondoftwo
 \fi
}%
\providecommand \@ifx [1]{%
 \ifx #1\expandafter \@firstoftwo
 \else \expandafter \@secondoftwo
 \fi
}%
\providecommand \natexlab [1]{#1}%
\providecommand \enquote  [1]{``#1''}%
\providecommand \bibnamefont  [1]{#1}%
\providecommand \bibfnamefont [1]{#1}%
\providecommand \citenamefont [1]{#1}%
\providecommand \href@noop [0]{\@secondoftwo}%
\providecommand \href [0]{\begingroup \@sanitize@url \@href}%
\providecommand \@href[1]{\@@startlink{#1}\@@href}%
\providecommand \@@href[1]{\endgroup#1\@@endlink}%
\providecommand \@sanitize@url [0]{\catcode `\\12\catcode `\$12\catcode
  `\&12\catcode `\#12\catcode `\^12\catcode `\_12\catcode `\%12\relax}%
\providecommand \@@startlink[1]{}%
\providecommand \@@endlink[0]{}%
\providecommand \url  [0]{\begingroup\@sanitize@url \@url }%
\providecommand \@url [1]{\endgroup\@href {#1}{\urlprefix }}%
\providecommand \urlprefix  [0]{URL }%
\providecommand \Eprint [0]{\href }%
\providecommand \doibase [0]{http://dx.doi.org/}%
\providecommand \selectlanguage [0]{\@gobble}%
\providecommand \bibinfo  [0]{\@secondoftwo}%
\providecommand \bibfield  [0]{\@secondoftwo}%
\providecommand \translation [1]{[#1]}%
\providecommand \BibitemOpen [0]{}%
\providecommand \bibitemStop [0]{}%
\providecommand \bibitemNoStop [0]{.\EOS\space}%
\providecommand \EOS [0]{\spacefactor3000\relax}%
\providecommand \BibitemShut  [1]{\csname bibitem#1\endcsname}%
\let\auto@bib@innerbib\@empty
\bibitem [{\citenamefont {Torres}(2019)}]{torres2019perspective}%
  \BibitemOpen
  \bibfield  {author} {\bibinfo {author} {\bibfnamefont {L.~E.~F.}\
  \bibnamefont {Torres}},\ }\href@noop {} {\bibfield  {journal} {\bibinfo
  {journal} {Journal of Physics: Materials}\ }\textbf {\bibinfo {volume} {3}},\
  \bibinfo {pages} {014002} (\bibinfo {year} {2019})}\BibitemShut {NoStop}%
\bibitem [{\citenamefont {Ghatak}\ and\ \citenamefont
  {Das}(2019)}]{ghatak2019new}%
  \BibitemOpen
  \bibfield  {author} {\bibinfo {author} {\bibfnamefont {A.}~\bibnamefont
  {Ghatak}}\ and\ \bibinfo {author} {\bibfnamefont {T.}~\bibnamefont {Das}},\
  }\href@noop {} {\bibfield  {journal} {\bibinfo  {journal} {Journal of
  Physics: Condensed Matter}\ }\textbf {\bibinfo {volume} {31}},\ \bibinfo
  {pages} {263001} (\bibinfo {year} {2019})}\BibitemShut {NoStop}%
\bibitem [{\citenamefont {Bergholtz}\ \emph {et~al.}(2019)\citenamefont
  {Bergholtz}, \citenamefont {Budich},\ and\ \citenamefont
  {Kunst}}]{bergholtz2019exceptional}%
  \BibitemOpen
  \bibfield  {author} {\bibinfo {author} {\bibfnamefont {E.~J.}\ \bibnamefont
  {Bergholtz}}, \bibinfo {author} {\bibfnamefont {J.~C.}\ \bibnamefont
  {Budich}}, \ and\ \bibinfo {author} {\bibfnamefont {F.~K.}\ \bibnamefont
  {Kunst}},\ }\href@noop {} {\bibfield  {journal} {\bibinfo  {journal} {arXiv
  preprint arXiv:1912.10048}\ } (\bibinfo {year} {2019})}\BibitemShut {NoStop}%
\bibitem [{\citenamefont {Gong}\ \emph {et~al.}(2018)\citenamefont {Gong},
  \citenamefont {Ashida}, \citenamefont {Kawabata}, \citenamefont {Takasan},
  \citenamefont {Higashikawa},\ and\ \citenamefont
  {Ueda}}]{gong2018topological}%
  \BibitemOpen
  \bibfield  {author} {\bibinfo {author} {\bibfnamefont {Z.}~\bibnamefont
  {Gong}}, \bibinfo {author} {\bibfnamefont {Y.}~\bibnamefont {Ashida}},
  \bibinfo {author} {\bibfnamefont {K.}~\bibnamefont {Kawabata}}, \bibinfo
  {author} {\bibfnamefont {K.}~\bibnamefont {Takasan}}, \bibinfo {author}
  {\bibfnamefont {S.}~\bibnamefont {Higashikawa}}, \ and\ \bibinfo {author}
  {\bibfnamefont {M.}~\bibnamefont {Ueda}},\ }\href@noop {} {\bibfield
  {journal} {\bibinfo  {journal} {Physical Review X}\ }\textbf {\bibinfo
  {volume} {8}},\ \bibinfo {pages} {031079} (\bibinfo {year}
  {2018})}\BibitemShut {NoStop}%
\bibitem [{\citenamefont {Xu}\ \emph {et~al.}(2017)\citenamefont {Xu},
  \citenamefont {Wang},\ and\ \citenamefont {Duan}}]{xu2017weyl}%
  \BibitemOpen
  \bibfield  {author} {\bibinfo {author} {\bibfnamefont {Y.}~\bibnamefont
  {Xu}}, \bibinfo {author} {\bibfnamefont {S.-T.}\ \bibnamefont {Wang}}, \ and\
  \bibinfo {author} {\bibfnamefont {L.-M.}\ \bibnamefont {Duan}},\ }\href@noop
  {} {\bibfield  {journal} {\bibinfo  {journal} {Physical review letters}\
  }\textbf {\bibinfo {volume} {118}},\ \bibinfo {pages} {045701} (\bibinfo
  {year} {2017})}\BibitemShut {NoStop}%
\bibitem [{\citenamefont {Goldman}\ \emph {et~al.}(2016)\citenamefont
  {Goldman}, \citenamefont {Budich},\ and\ \citenamefont
  {Zoller}}]{goldman2016topological}%
  \BibitemOpen
  \bibfield  {author} {\bibinfo {author} {\bibfnamefont {N.}~\bibnamefont
  {Goldman}}, \bibinfo {author} {\bibfnamefont {J.~C.}\ \bibnamefont {Budich}},
  \ and\ \bibinfo {author} {\bibfnamefont {P.}~\bibnamefont {Zoller}},\
  }\href@noop {} {\bibfield  {journal} {\bibinfo  {journal} {Nature Physics}\
  }\textbf {\bibinfo {volume} {12}},\ \bibinfo {pages} {639} (\bibinfo {year}
  {2016})}\BibitemShut {NoStop}%
\bibitem [{\citenamefont {Ozawa}\ \emph {et~al.}(2019)\citenamefont {Ozawa},
  \citenamefont {Price}, \citenamefont {Amo}, \citenamefont {Goldman},
  \citenamefont {Hafezi}, \citenamefont {Lu}, \citenamefont {Rechtsman},
  \citenamefont {Schuster}, \citenamefont {Simon}, \citenamefont {Zilberberg}
  \emph {et~al.}}]{ozawa2019topological}%
  \BibitemOpen
  \bibfield  {author} {\bibinfo {author} {\bibfnamefont {T.}~\bibnamefont
  {Ozawa}}, \bibinfo {author} {\bibfnamefont {H.~M.}\ \bibnamefont {Price}},
  \bibinfo {author} {\bibfnamefont {A.}~\bibnamefont {Amo}}, \bibinfo {author}
  {\bibfnamefont {N.}~\bibnamefont {Goldman}}, \bibinfo {author} {\bibfnamefont
  {M.}~\bibnamefont {Hafezi}}, \bibinfo {author} {\bibfnamefont
  {L.}~\bibnamefont {Lu}}, \bibinfo {author} {\bibfnamefont {M.~C.}\
  \bibnamefont {Rechtsman}}, \bibinfo {author} {\bibfnamefont {D.}~\bibnamefont
  {Schuster}}, \bibinfo {author} {\bibfnamefont {J.}~\bibnamefont {Simon}},
  \bibinfo {author} {\bibfnamefont {O.}~\bibnamefont {Zilberberg}},  \emph
  {et~al.},\ }\href@noop {} {\bibfield  {journal} {\bibinfo  {journal} {Reviews
  of Modern Physics}\ }\textbf {\bibinfo {volume} {91}},\ \bibinfo {pages}
  {015006} (\bibinfo {year} {2019})}\BibitemShut {NoStop}%
\bibitem [{\citenamefont {Noh}\ \emph {et~al.}(2018)\citenamefont {Noh},
  \citenamefont {Benalcazar}, \citenamefont {Huang}, \citenamefont {Collins},
  \citenamefont {Chen}, \citenamefont {Hughes},\ and\ \citenamefont
  {Rechtsman}}]{noh2018topological}%
  \BibitemOpen
  \bibfield  {author} {\bibinfo {author} {\bibfnamefont {J.}~\bibnamefont
  {Noh}}, \bibinfo {author} {\bibfnamefont {W.~A.}\ \bibnamefont {Benalcazar}},
  \bibinfo {author} {\bibfnamefont {S.}~\bibnamefont {Huang}}, \bibinfo
  {author} {\bibfnamefont {M.~J.}\ \bibnamefont {Collins}}, \bibinfo {author}
  {\bibfnamefont {K.~P.}\ \bibnamefont {Chen}}, \bibinfo {author}
  {\bibfnamefont {T.~L.}\ \bibnamefont {Hughes}}, \ and\ \bibinfo {author}
  {\bibfnamefont {M.~C.}\ \bibnamefont {Rechtsman}},\ }\href@noop {} {\bibfield
   {journal} {\bibinfo  {journal} {Nature Photonics}\ }\textbf {\bibinfo
  {volume} {12}},\ \bibinfo {pages} {408} (\bibinfo {year} {2018})}\BibitemShut
  {NoStop}%
\bibitem [{\citenamefont {Poli}\ \emph {et~al.}(2015)\citenamefont {Poli},
  \citenamefont {Bellec}, \citenamefont {Kuhl}, \citenamefont {Mortessagne},\
  and\ \citenamefont {Schomerus}}]{poli2015selective}%
  \BibitemOpen
  \bibfield  {author} {\bibinfo {author} {\bibfnamefont {C.}~\bibnamefont
  {Poli}}, \bibinfo {author} {\bibfnamefont {M.}~\bibnamefont {Bellec}},
  \bibinfo {author} {\bibfnamefont {U.}~\bibnamefont {Kuhl}}, \bibinfo {author}
  {\bibfnamefont {F.}~\bibnamefont {Mortessagne}}, \ and\ \bibinfo {author}
  {\bibfnamefont {H.}~\bibnamefont {Schomerus}},\ }\href@noop {} {\bibfield
  {journal} {\bibinfo  {journal} {Nature communications}\ }\textbf {\bibinfo
  {volume} {6}},\ \bibinfo {pages} {1} (\bibinfo {year} {2015})}\BibitemShut
  {NoStop}%
\bibitem [{\citenamefont {Cao}\ and\ \citenamefont
  {Wiersig}(2015)}]{cao2015dielectric}%
  \BibitemOpen
  \bibfield  {author} {\bibinfo {author} {\bibfnamefont {H.}~\bibnamefont
  {Cao}}\ and\ \bibinfo {author} {\bibfnamefont {J.}~\bibnamefont {Wiersig}},\
  }\href@noop {} {\bibfield  {journal} {\bibinfo  {journal} {Reviews of Modern
  Physics}\ }\textbf {\bibinfo {volume} {87}},\ \bibinfo {pages} {61} (\bibinfo
  {year} {2015})}\BibitemShut {NoStop}%
\bibitem [{\citenamefont {Bandres}\ \emph {et~al.}(2018)\citenamefont
  {Bandres}, \citenamefont {Wittek}, \citenamefont {Harari}, \citenamefont
  {Parto}, \citenamefont {Ren}, \citenamefont {Segev}, \citenamefont
  {Christodoulides},\ and\ \citenamefont
  {Khajavikhan}}]{bandres2018topological}%
  \BibitemOpen
  \bibfield  {author} {\bibinfo {author} {\bibfnamefont {M.~A.}\ \bibnamefont
  {Bandres}}, \bibinfo {author} {\bibfnamefont {S.}~\bibnamefont {Wittek}},
  \bibinfo {author} {\bibfnamefont {G.}~\bibnamefont {Harari}}, \bibinfo
  {author} {\bibfnamefont {M.}~\bibnamefont {Parto}}, \bibinfo {author}
  {\bibfnamefont {J.}~\bibnamefont {Ren}}, \bibinfo {author} {\bibfnamefont
  {M.}~\bibnamefont {Segev}}, \bibinfo {author} {\bibfnamefont {D.~N.}\
  \bibnamefont {Christodoulides}}, \ and\ \bibinfo {author} {\bibfnamefont
  {M.}~\bibnamefont {Khajavikhan}},\ }\href@noop {} {\bibfield  {journal}
  {\bibinfo  {journal} {Science}\ }\textbf {\bibinfo {volume} {359}},\ \bibinfo
  {pages} {eaar4005} (\bibinfo {year} {2018})}\BibitemShut {NoStop}%
\bibitem [{\citenamefont {Harari}\ \emph {et~al.}(2018)\citenamefont {Harari},
  \citenamefont {Bandres}, \citenamefont {Lumer}, \citenamefont {Rechtsman},
  \citenamefont {Chong}, \citenamefont {Khajavikhan}, \citenamefont
  {Christodoulides},\ and\ \citenamefont {Segev}}]{harari2018topological}%
  \BibitemOpen
  \bibfield  {author} {\bibinfo {author} {\bibfnamefont {G.}~\bibnamefont
  {Harari}}, \bibinfo {author} {\bibfnamefont {M.~A.}\ \bibnamefont {Bandres}},
  \bibinfo {author} {\bibfnamefont {Y.}~\bibnamefont {Lumer}}, \bibinfo
  {author} {\bibfnamefont {M.~C.}\ \bibnamefont {Rechtsman}}, \bibinfo {author}
  {\bibfnamefont {Y.~D.}\ \bibnamefont {Chong}}, \bibinfo {author}
  {\bibfnamefont {M.}~\bibnamefont {Khajavikhan}}, \bibinfo {author}
  {\bibfnamefont {D.~N.}\ \bibnamefont {Christodoulides}}, \ and\ \bibinfo
  {author} {\bibfnamefont {M.}~\bibnamefont {Segev}},\ }\href@noop {}
  {\bibfield  {journal} {\bibinfo  {journal} {Science}\ }\textbf {\bibinfo
  {volume} {359}},\ \bibinfo {pages} {eaar4003} (\bibinfo {year}
  {2018})}\BibitemShut {NoStop}%
\bibitem [{\citenamefont {Feng}\ \emph {et~al.}(2014)\citenamefont {Feng},
  \citenamefont {Wong}, \citenamefont {Ma}, \citenamefont {Wang},\ and\
  \citenamefont {Zhang}}]{feng2014single}%
  \BibitemOpen
  \bibfield  {author} {\bibinfo {author} {\bibfnamefont {L.}~\bibnamefont
  {Feng}}, \bibinfo {author} {\bibfnamefont {Z.~J.}\ \bibnamefont {Wong}},
  \bibinfo {author} {\bibfnamefont {R.-M.}\ \bibnamefont {Ma}}, \bibinfo
  {author} {\bibfnamefont {Y.}~\bibnamefont {Wang}}, \ and\ \bibinfo {author}
  {\bibfnamefont {X.}~\bibnamefont {Zhang}},\ }\href@noop {} {\bibfield
  {journal} {\bibinfo  {journal} {Science}\ }\textbf {\bibinfo {volume}
  {346}},\ \bibinfo {pages} {972} (\bibinfo {year} {2014})}\BibitemShut
  {NoStop}%
\bibitem [{\citenamefont {Kawabata}\ \emph {et~al.}(2019)\citenamefont
  {Kawabata}, \citenamefont {Bessho},\ and\ \citenamefont
  {Sato}}]{kawabata2019classification}%
  \BibitemOpen
  \bibfield  {author} {\bibinfo {author} {\bibfnamefont {K.}~\bibnamefont
  {Kawabata}}, \bibinfo {author} {\bibfnamefont {T.}~\bibnamefont {Bessho}}, \
  and\ \bibinfo {author} {\bibfnamefont {M.}~\bibnamefont {Sato}},\ }\href@noop
  {} {\bibfield  {journal} {\bibinfo  {journal} {Physical review letters}\
  }\textbf {\bibinfo {volume} {123}},\ \bibinfo {pages} {066405} (\bibinfo
  {year} {2019})}\BibitemShut {NoStop}%
\bibitem [{\citenamefont {Carlstr{\"o}m}\ \emph {et~al.}(2019)\citenamefont
  {Carlstr{\"o}m}, \citenamefont {St{\aa}lhammar}, \citenamefont {Budich},\
  and\ \citenamefont {Bergholtz}}]{carlstrom2019knotted}%
  \BibitemOpen
  \bibfield  {author} {\bibinfo {author} {\bibfnamefont {J.}~\bibnamefont
  {Carlstr{\"o}m}}, \bibinfo {author} {\bibfnamefont {M.}~\bibnamefont
  {St{\aa}lhammar}}, \bibinfo {author} {\bibfnamefont {J.~C.}\ \bibnamefont
  {Budich}}, \ and\ \bibinfo {author} {\bibfnamefont {E.~J.}\ \bibnamefont
  {Bergholtz}},\ }\href@noop {} {\bibfield  {journal} {\bibinfo  {journal}
  {Physical Review B}\ }\textbf {\bibinfo {volume} {99}},\ \bibinfo {pages}
  {161115} (\bibinfo {year} {2019})}\BibitemShut {NoStop}%
\bibitem [{\citenamefont {Wang}\ \emph {et~al.}(2018)\citenamefont {Wang},
  \citenamefont {Ruan},\ and\ \citenamefont {Zhang}}]{wang2018non}%
  \BibitemOpen
  \bibfield  {author} {\bibinfo {author} {\bibfnamefont {H.}~\bibnamefont
  {Wang}}, \bibinfo {author} {\bibfnamefont {J.}~\bibnamefont {Ruan}}, \ and\
  \bibinfo {author} {\bibfnamefont {H.}~\bibnamefont {Zhang}},\ }\href@noop {}
  {\bibfield  {journal} {\bibinfo  {journal} {arXiv preprint arXiv:1808.06162}\
  } (\bibinfo {year} {2018})}\BibitemShut {NoStop}%
\bibitem [{\citenamefont {Budich}\ \emph {et~al.}(2019)\citenamefont {Budich},
  \citenamefont {Carlstr{\"o}m}, \citenamefont {Kunst},\ and\ \citenamefont
  {Bergholtz}}]{budich2019symmetry}%
  \BibitemOpen
  \bibfield  {author} {\bibinfo {author} {\bibfnamefont {J.~C.}\ \bibnamefont
  {Budich}}, \bibinfo {author} {\bibfnamefont {J.}~\bibnamefont
  {Carlstr{\"o}m}}, \bibinfo {author} {\bibfnamefont {F.~K.}\ \bibnamefont
  {Kunst}}, \ and\ \bibinfo {author} {\bibfnamefont {E.~J.}\ \bibnamefont
  {Bergholtz}},\ }\href@noop {} {\bibfield  {journal} {\bibinfo  {journal}
  {Physical Review B}\ }\textbf {\bibinfo {volume} {99}},\ \bibinfo {pages}
  {041406} (\bibinfo {year} {2019})}\BibitemShut {NoStop}%
\bibitem [{\citenamefont {Wang}\ \emph {et~al.}(2019)\citenamefont {Wang},
  \citenamefont {Ruan},\ and\ \citenamefont {Zhang}}]{wang2019non}%
  \BibitemOpen
  \bibfield  {author} {\bibinfo {author} {\bibfnamefont {H.}~\bibnamefont
  {Wang}}, \bibinfo {author} {\bibfnamefont {J.}~\bibnamefont {Ruan}}, \ and\
  \bibinfo {author} {\bibfnamefont {H.}~\bibnamefont {Zhang}},\ }\href@noop {}
  {\bibfield  {journal} {\bibinfo  {journal} {Physical Review B}\ }\textbf
  {\bibinfo {volume} {99}},\ \bibinfo {pages} {075130} (\bibinfo {year}
  {2019})}\BibitemShut {NoStop}%
\bibitem [{\citenamefont {Yang}\ and\ \citenamefont {Hu}(2019)}]{yang2019non}%
  \BibitemOpen
  \bibfield  {author} {\bibinfo {author} {\bibfnamefont {Z.}~\bibnamefont
  {Yang}}\ and\ \bibinfo {author} {\bibfnamefont {J.}~\bibnamefont {Hu}},\
  }\href@noop {} {\bibfield  {journal} {\bibinfo  {journal} {Physical Review
  B}\ }\textbf {\bibinfo {volume} {99}},\ \bibinfo {pages} {081102} (\bibinfo
  {year} {2019})}\BibitemShut {NoStop}%
\bibitem [{\citenamefont {Rui}\ \emph {et~al.}(2019)\citenamefont {Rui},
  \citenamefont {Zhao},\ and\ \citenamefont {Schnyder}}]{rui2019topology}%
  \BibitemOpen
  \bibfield  {author} {\bibinfo {author} {\bibfnamefont {W.}~\bibnamefont
  {Rui}}, \bibinfo {author} {\bibfnamefont {Y.}~\bibnamefont {Zhao}}, \ and\
  \bibinfo {author} {\bibfnamefont {A.~P.}\ \bibnamefont {Schnyder}},\
  }\href@noop {} {\bibfield  {journal} {\bibinfo  {journal} {Physical Review
  B}\ }\textbf {\bibinfo {volume} {99}},\ \bibinfo {pages} {241110} (\bibinfo
  {year} {2019})}\BibitemShut {NoStop}%
\bibitem [{\citenamefont {Papaj}\ \emph {et~al.}(2019)\citenamefont {Papaj},
  \citenamefont {Isobe},\ and\ \citenamefont {Fu}}]{papaj2019nodal}%
  \BibitemOpen
  \bibfield  {author} {\bibinfo {author} {\bibfnamefont {M.}~\bibnamefont
  {Papaj}}, \bibinfo {author} {\bibfnamefont {H.}~\bibnamefont {Isobe}}, \ and\
  \bibinfo {author} {\bibfnamefont {L.}~\bibnamefont {Fu}},\ }\href@noop {}
  {\bibfield  {journal} {\bibinfo  {journal} {Physical Review B}\ }\textbf
  {\bibinfo {volume} {99}},\ \bibinfo {pages} {201107} (\bibinfo {year}
  {2019})}\BibitemShut {NoStop}%
\bibitem [{\citenamefont {Banerjee}\ and\ \citenamefont
  {Narayan}(2020)}]{banerjee2020non}%
  \BibitemOpen
  \bibfield  {author} {\bibinfo {author} {\bibfnamefont {A.}~\bibnamefont
  {Banerjee}}\ and\ \bibinfo {author} {\bibfnamefont {A.}~\bibnamefont
  {Narayan}},\ }\href@noop {} {\bibfield  {journal} {\bibinfo  {journal} {arXiv
  preprint arXiv:2001.11188}\ } (\bibinfo {year} {2020})}\BibitemShut {NoStop}%
\bibitem [{\citenamefont {Heiss}(2012)}]{heiss2012physics}%
  \BibitemOpen
  \bibfield  {author} {\bibinfo {author} {\bibfnamefont {W.}~\bibnamefont
  {Heiss}},\ }\href@noop {} {\bibfield  {journal} {\bibinfo  {journal} {Journal
  of Physics A: Mathematical and Theoretical}\ }\textbf {\bibinfo {volume}
  {45}},\ \bibinfo {pages} {444016} (\bibinfo {year} {2012})}\BibitemShut
  {NoStop}%
\bibitem [{\citenamefont {Alvarez}\ \emph
  {et~al.}(2018{\natexlab{a}})\citenamefont {Alvarez}, \citenamefont {Vargas},\
  and\ \citenamefont {Torres}}]{alvarez2018non}%
  \BibitemOpen
  \bibfield  {author} {\bibinfo {author} {\bibfnamefont {V.~M.}\ \bibnamefont
  {Alvarez}}, \bibinfo {author} {\bibfnamefont {J.~B.}\ \bibnamefont {Vargas}},
  \ and\ \bibinfo {author} {\bibfnamefont {L.~F.}\ \bibnamefont {Torres}},\
  }\href@noop {} {\bibfield  {journal} {\bibinfo  {journal} {Physical Review
  B}\ }\textbf {\bibinfo {volume} {97}},\ \bibinfo {pages} {121401} (\bibinfo
  {year} {2018}{\natexlab{a}})}\BibitemShut {NoStop}%
\bibitem [{\citenamefont {Heiss}(2016)}]{heiss2016mathematical}%
  \BibitemOpen
  \bibfield  {author} {\bibinfo {author} {\bibfnamefont {D.}~\bibnamefont
  {Heiss}},\ }\href@noop {} {\bibfield  {journal} {\bibinfo  {journal} {Nature
  Physics}\ }\textbf {\bibinfo {volume} {12}},\ \bibinfo {pages} {823}
  (\bibinfo {year} {2016})}\BibitemShut {NoStop}%
\bibitem [{\citenamefont {Oka}\ and\ \citenamefont
  {Aoki}(2009)}]{oka2009photovoltaic}%
  \BibitemOpen
  \bibfield  {author} {\bibinfo {author} {\bibfnamefont {T.}~\bibnamefont
  {Oka}}\ and\ \bibinfo {author} {\bibfnamefont {H.}~\bibnamefont {Aoki}},\
  }\href@noop {} {\bibfield  {journal} {\bibinfo  {journal} {Physical Review
  B}\ }\textbf {\bibinfo {volume} {79}},\ \bibinfo {pages} {081406} (\bibinfo
  {year} {2009})}\BibitemShut {NoStop}%
\bibitem [{\citenamefont {Lindner}\ \emph {et~al.}(2011)\citenamefont
  {Lindner}, \citenamefont {Refael},\ and\ \citenamefont
  {Galitski}}]{lindner2011floquet}%
  \BibitemOpen
  \bibfield  {author} {\bibinfo {author} {\bibfnamefont {N.~H.}\ \bibnamefont
  {Lindner}}, \bibinfo {author} {\bibfnamefont {G.}~\bibnamefont {Refael}}, \
  and\ \bibinfo {author} {\bibfnamefont {V.}~\bibnamefont {Galitski}},\
  }\href@noop {} {\bibfield  {journal} {\bibinfo  {journal} {Nature Physics}\
  }\textbf {\bibinfo {volume} {7}},\ \bibinfo {pages} {490} (\bibinfo {year}
  {2011})}\BibitemShut {NoStop}%
\bibitem [{\citenamefont {Cayssol}\ \emph {et~al.}(2013)\citenamefont
  {Cayssol}, \citenamefont {D{\'o}ra}, \citenamefont {Simon},\ and\
  \citenamefont {Moessner}}]{cayssol2013floquet}%
  \BibitemOpen
  \bibfield  {author} {\bibinfo {author} {\bibfnamefont {J.}~\bibnamefont
  {Cayssol}}, \bibinfo {author} {\bibfnamefont {B.}~\bibnamefont {D{\'o}ra}},
  \bibinfo {author} {\bibfnamefont {F.}~\bibnamefont {Simon}}, \ and\ \bibinfo
  {author} {\bibfnamefont {R.}~\bibnamefont {Moessner}},\ }\href@noop {}
  {\bibfield  {journal} {\bibinfo  {journal} {physica status solidi
  (RRL)--Rapid Research Letters}\ }\textbf {\bibinfo {volume} {7}},\ \bibinfo
  {pages} {101} (\bibinfo {year} {2013})}\BibitemShut {NoStop}%
\bibitem [{\citenamefont {Wang}\ \emph
  {et~al.}(2013{\natexlab{a}})\citenamefont {Wang}, \citenamefont {Steinberg},
  \citenamefont {Jarillo-Herrero},\ and\ \citenamefont
  {Gedik}}]{wang2013observation}%
  \BibitemOpen
  \bibfield  {author} {\bibinfo {author} {\bibfnamefont {Y.}~\bibnamefont
  {Wang}}, \bibinfo {author} {\bibfnamefont {H.}~\bibnamefont {Steinberg}},
  \bibinfo {author} {\bibfnamefont {P.}~\bibnamefont {Jarillo-Herrero}}, \ and\
  \bibinfo {author} {\bibfnamefont {N.}~\bibnamefont {Gedik}},\ }\href@noop {}
  {\bibfield  {journal} {\bibinfo  {journal} {Science}\ }\textbf {\bibinfo
  {volume} {342}},\ \bibinfo {pages} {453} (\bibinfo {year}
  {2013}{\natexlab{a}})}\BibitemShut {NoStop}%
\bibitem [{\citenamefont {Rechtsman}\ \emph {et~al.}(2013)\citenamefont
  {Rechtsman}, \citenamefont {Zeuner}, \citenamefont {Plotnik}, \citenamefont
  {Lumer}, \citenamefont {Podolsky}, \citenamefont {Dreisow}, \citenamefont
  {Nolte}, \citenamefont {Segev},\ and\ \citenamefont
  {Szameit}}]{rechtsman2013photonic}%
  \BibitemOpen
  \bibfield  {author} {\bibinfo {author} {\bibfnamefont {M.~C.}\ \bibnamefont
  {Rechtsman}}, \bibinfo {author} {\bibfnamefont {J.~M.}\ \bibnamefont
  {Zeuner}}, \bibinfo {author} {\bibfnamefont {Y.}~\bibnamefont {Plotnik}},
  \bibinfo {author} {\bibfnamefont {Y.}~\bibnamefont {Lumer}}, \bibinfo
  {author} {\bibfnamefont {D.}~\bibnamefont {Podolsky}}, \bibinfo {author}
  {\bibfnamefont {F.}~\bibnamefont {Dreisow}}, \bibinfo {author} {\bibfnamefont
  {S.}~\bibnamefont {Nolte}}, \bibinfo {author} {\bibfnamefont
  {M.}~\bibnamefont {Segev}}, \ and\ \bibinfo {author} {\bibfnamefont
  {A.}~\bibnamefont {Szameit}},\ }\href@noop {} {\bibfield  {journal} {\bibinfo
   {journal} {Nature}\ }\textbf {\bibinfo {volume} {496}},\ \bibinfo {pages}
  {196} (\bibinfo {year} {2013})}\BibitemShut {NoStop}%
\bibitem [{\citenamefont {Rudner}\ and\ \citenamefont
  {Lindner}(2020)}]{rudner2020floquet}%
  \BibitemOpen
  \bibfield  {author} {\bibinfo {author} {\bibfnamefont {M.~S.}\ \bibnamefont
  {Rudner}}\ and\ \bibinfo {author} {\bibfnamefont {N.~H.}\ \bibnamefont
  {Lindner}},\ }\href@noop {} {\bibfield  {journal} {\bibinfo  {journal} {arXiv
  preprint arXiv:2003.08252}\ } (\bibinfo {year} {2020})}\BibitemShut {NoStop}%
\bibitem [{\citenamefont {Kitagawa}\ \emph {et~al.}(2011)\citenamefont
  {Kitagawa}, \citenamefont {Oka}, \citenamefont {Brataas}, \citenamefont
  {Fu},\ and\ \citenamefont {Demler}}]{kitagawa2011transport}%
  \BibitemOpen
  \bibfield  {author} {\bibinfo {author} {\bibfnamefont {T.}~\bibnamefont
  {Kitagawa}}, \bibinfo {author} {\bibfnamefont {T.}~\bibnamefont {Oka}},
  \bibinfo {author} {\bibfnamefont {A.}~\bibnamefont {Brataas}}, \bibinfo
  {author} {\bibfnamefont {L.}~\bibnamefont {Fu}}, \ and\ \bibinfo {author}
  {\bibfnamefont {E.}~\bibnamefont {Demler}},\ }\href@noop {} {\bibfield
  {journal} {\bibinfo  {journal} {Physical Review B}\ }\textbf {\bibinfo
  {volume} {84}},\ \bibinfo {pages} {235108} (\bibinfo {year}
  {2011})}\BibitemShut {NoStop}%
\bibitem [{\citenamefont {Gu}\ \emph {et~al.}(2011)\citenamefont {Gu},
  \citenamefont {Fertig}, \citenamefont {Arovas},\ and\ \citenamefont
  {Auerbach}}]{gu2011floquet}%
  \BibitemOpen
  \bibfield  {author} {\bibinfo {author} {\bibfnamefont {Z.}~\bibnamefont
  {Gu}}, \bibinfo {author} {\bibfnamefont {H.}~\bibnamefont {Fertig}}, \bibinfo
  {author} {\bibfnamefont {D.~P.}\ \bibnamefont {Arovas}}, \ and\ \bibinfo
  {author} {\bibfnamefont {A.}~\bibnamefont {Auerbach}},\ }\href@noop {}
  {\bibfield  {journal} {\bibinfo  {journal} {Physical review letters}\
  }\textbf {\bibinfo {volume} {107}},\ \bibinfo {pages} {216601} (\bibinfo
  {year} {2011})}\BibitemShut {NoStop}%
\bibitem [{\citenamefont {Ezawa}(2013)}]{ezawa2013photoinduced}%
  \BibitemOpen
  \bibfield  {author} {\bibinfo {author} {\bibfnamefont {M.}~\bibnamefont
  {Ezawa}},\ }\href@noop {} {\bibfield  {journal} {\bibinfo  {journal}
  {Physical review letters}\ }\textbf {\bibinfo {volume} {110}},\ \bibinfo
  {pages} {026603} (\bibinfo {year} {2013})}\BibitemShut {NoStop}%
\bibitem [{\citenamefont {Perez-Piskunow}\ \emph {et~al.}(2014)\citenamefont
  {Perez-Piskunow}, \citenamefont {Usaj}, \citenamefont {Balseiro},\ and\
  \citenamefont {Torres}}]{perez2014floquet}%
  \BibitemOpen
  \bibfield  {author} {\bibinfo {author} {\bibfnamefont {P.~M.}\ \bibnamefont
  {Perez-Piskunow}}, \bibinfo {author} {\bibfnamefont {G.}~\bibnamefont
  {Usaj}}, \bibinfo {author} {\bibfnamefont {C.~A.}\ \bibnamefont {Balseiro}},
  \ and\ \bibinfo {author} {\bibfnamefont {L.~F.}\ \bibnamefont {Torres}},\
  }\href@noop {} {\bibfield  {journal} {\bibinfo  {journal} {Physical Review
  B}\ }\textbf {\bibinfo {volume} {89}},\ \bibinfo {pages} {121401} (\bibinfo
  {year} {2014})}\BibitemShut {NoStop}%
\bibitem [{\citenamefont {Wang}\ \emph {et~al.}(2014)\citenamefont {Wang},
  \citenamefont {Wang}, \citenamefont {Shen}, \citenamefont {Sheng},\ and\
  \citenamefont {Xing}}]{wang2014floquet}%
  \BibitemOpen
  \bibfield  {author} {\bibinfo {author} {\bibfnamefont {R.}~\bibnamefont
  {Wang}}, \bibinfo {author} {\bibfnamefont {B.}~\bibnamefont {Wang}}, \bibinfo
  {author} {\bibfnamefont {R.}~\bibnamefont {Shen}}, \bibinfo {author}
  {\bibfnamefont {L.}~\bibnamefont {Sheng}}, \ and\ \bibinfo {author}
  {\bibfnamefont {D.}~\bibnamefont {Xing}},\ }\href@noop {} {\bibfield
  {journal} {\bibinfo  {journal} {EPL (Europhysics Letters)}\ }\textbf
  {\bibinfo {volume} {105}},\ \bibinfo {pages} {17004} (\bibinfo {year}
  {2014})}\BibitemShut {NoStop}%
\bibitem [{\citenamefont {Gonz{\'a}lez}\ and\ \citenamefont
  {Molina}(2016)}]{gonzalez2016macroscopic}%
  \BibitemOpen
  \bibfield  {author} {\bibinfo {author} {\bibfnamefont {J.}~\bibnamefont
  {Gonz{\'a}lez}}\ and\ \bibinfo {author} {\bibfnamefont {R.~A.}\ \bibnamefont
  {Molina}},\ }\href@noop {} {\bibfield  {journal} {\bibinfo  {journal}
  {Physical review letters}\ }\textbf {\bibinfo {volume} {116}},\ \bibinfo
  {pages} {156803} (\bibinfo {year} {2016})}\BibitemShut {NoStop}%
\bibitem [{\citenamefont {Narayan}(2016)}]{narayan2016tunable}%
  \BibitemOpen
  \bibfield  {author} {\bibinfo {author} {\bibfnamefont {A.}~\bibnamefont
  {Narayan}},\ }\href@noop {} {\bibfield  {journal} {\bibinfo  {journal}
  {Physical Review B}\ }\textbf {\bibinfo {volume} {94}},\ \bibinfo {pages}
  {041409} (\bibinfo {year} {2016})}\BibitemShut {NoStop}%
\bibitem [{\citenamefont {Yan}\ and\ \citenamefont
  {Wang}(2016)}]{yan2016tunable}%
  \BibitemOpen
  \bibfield  {author} {\bibinfo {author} {\bibfnamefont {Z.}~\bibnamefont
  {Yan}}\ and\ \bibinfo {author} {\bibfnamefont {Z.}~\bibnamefont {Wang}},\
  }\href@noop {} {\bibfield  {journal} {\bibinfo  {journal} {Physical review
  letters}\ }\textbf {\bibinfo {volume} {117}},\ \bibinfo {pages} {087402}
  (\bibinfo {year} {2016})}\BibitemShut {NoStop}%
\bibitem [{\citenamefont {Taguchi}\ \emph {et~al.}(2016)\citenamefont
  {Taguchi}, \citenamefont {Xu}, \citenamefont {Yamakage},\ and\ \citenamefont
  {Law}}]{taguchi2016photovoltaic}%
  \BibitemOpen
  \bibfield  {author} {\bibinfo {author} {\bibfnamefont {K.}~\bibnamefont
  {Taguchi}}, \bibinfo {author} {\bibfnamefont {D.-H.}\ \bibnamefont {Xu}},
  \bibinfo {author} {\bibfnamefont {A.}~\bibnamefont {Yamakage}}, \ and\
  \bibinfo {author} {\bibfnamefont {K.}~\bibnamefont {Law}},\ }\href@noop {}
  {\bibfield  {journal} {\bibinfo  {journal} {Physical Review B}\ }\textbf
  {\bibinfo {volume} {94}},\ \bibinfo {pages} {155206} (\bibinfo {year}
  {2016})}\BibitemShut {NoStop}%
\bibitem [{\citenamefont {Chan}\ \emph {et~al.}(2016)\citenamefont {Chan},
  \citenamefont {Oh}, \citenamefont {Han},\ and\ \citenamefont
  {Lee}}]{chan2016type}%
  \BibitemOpen
  \bibfield  {author} {\bibinfo {author} {\bibfnamefont {C.-K.}\ \bibnamefont
  {Chan}}, \bibinfo {author} {\bibfnamefont {Y.-T.}\ \bibnamefont {Oh}},
  \bibinfo {author} {\bibfnamefont {J.~H.}\ \bibnamefont {Han}}, \ and\
  \bibinfo {author} {\bibfnamefont {P.~A.}\ \bibnamefont {Lee}},\ }\href@noop
  {} {\bibfield  {journal} {\bibinfo  {journal} {Physical Review B}\ }\textbf
  {\bibinfo {volume} {94}},\ \bibinfo {pages} {121106} (\bibinfo {year}
  {2016})}\BibitemShut {NoStop}%
\bibitem [{\citenamefont {Inoue}\ and\ \citenamefont
  {Tanaka}(2010)}]{inoue2010photoinduced}%
  \BibitemOpen
  \bibfield  {author} {\bibinfo {author} {\bibfnamefont {J.-i.}\ \bibnamefont
  {Inoue}}\ and\ \bibinfo {author} {\bibfnamefont {A.}~\bibnamefont {Tanaka}},\
  }\href@noop {} {\bibfield  {journal} {\bibinfo  {journal} {Physical review
  letters}\ }\textbf {\bibinfo {volume} {105}},\ \bibinfo {pages} {017401}
  (\bibinfo {year} {2010})}\BibitemShut {NoStop}%
\bibitem [{\citenamefont {Saha}(2016)}]{saha2016photoinduced}%
  \BibitemOpen
  \bibfield  {author} {\bibinfo {author} {\bibfnamefont {K.}~\bibnamefont
  {Saha}},\ }\href@noop {} {\bibfield  {journal} {\bibinfo  {journal} {Physical
  Review B}\ }\textbf {\bibinfo {volume} {94}},\ \bibinfo {pages} {081103}
  (\bibinfo {year} {2016})}\BibitemShut {NoStop}%
\bibitem [{\citenamefont {Narayan}(2015)}]{narayan2015floquet}%
  \BibitemOpen
  \bibfield  {author} {\bibinfo {author} {\bibfnamefont {A.}~\bibnamefont
  {Narayan}},\ }\href@noop {} {\bibfield  {journal} {\bibinfo  {journal}
  {Physical Review B}\ }\textbf {\bibinfo {volume} {91}},\ \bibinfo {pages}
  {205445} (\bibinfo {year} {2015})}\BibitemShut {NoStop}%
\bibitem [{\citenamefont {Zhang}\ and\ \citenamefont
  {Gong}(2020)}]{zhang2020non}%
  \BibitemOpen
  \bibfield  {author} {\bibinfo {author} {\bibfnamefont {X.}~\bibnamefont
  {Zhang}}\ and\ \bibinfo {author} {\bibfnamefont {J.}~\bibnamefont {Gong}},\
  }\href@noop {} {\bibfield  {journal} {\bibinfo  {journal} {Physical Review
  B}\ }\textbf {\bibinfo {volume} {101}},\ \bibinfo {pages} {045415} (\bibinfo
  {year} {2020})}\BibitemShut {NoStop}%
\bibitem [{\citenamefont {Zhou}\ and\ \citenamefont
  {Gong}(2018)}]{zhou2018non}%
  \BibitemOpen
  \bibfield  {author} {\bibinfo {author} {\bibfnamefont {L.}~\bibnamefont
  {Zhou}}\ and\ \bibinfo {author} {\bibfnamefont {J.}~\bibnamefont {Gong}},\
  }\href@noop {} {\bibfield  {journal} {\bibinfo  {journal} {Physical Review
  B}\ }\textbf {\bibinfo {volume} {98}},\ \bibinfo {pages} {205417} (\bibinfo
  {year} {2018})}\BibitemShut {NoStop}%
\bibitem [{\citenamefont {Zhou}(2019)}]{zhou2019dynamical}%
  \BibitemOpen
  \bibfield  {author} {\bibinfo {author} {\bibfnamefont {L.}~\bibnamefont
  {Zhou}},\ }\href@noop {} {\bibfield  {journal} {\bibinfo  {journal} {Physical
  Review B}\ }\textbf {\bibinfo {volume} {100}},\ \bibinfo {pages} {184314}
  (\bibinfo {year} {2019})}\BibitemShut {NoStop}%
\bibitem [{\citenamefont {Wu}\ and\ \citenamefont {An}(2020)}]{wu2020floquet}%
  \BibitemOpen
  \bibfield  {author} {\bibinfo {author} {\bibfnamefont {H.}~\bibnamefont
  {Wu}}\ and\ \bibinfo {author} {\bibfnamefont {J.-H.}\ \bibnamefont {An}},\
  }\href@noop {} {\bibfield  {journal} {\bibinfo  {journal} {arXiv preprint
  arXiv:2003.08055}\ } (\bibinfo {year} {2020})}\BibitemShut {NoStop}%
\bibitem [{\citenamefont {H{\"o}ckendorf}\ \emph {et~al.}(2019)\citenamefont
  {H{\"o}ckendorf}, \citenamefont {Alvermann},\ and\ \citenamefont
  {Fehske}}]{hockendorf2019non}%
  \BibitemOpen
  \bibfield  {author} {\bibinfo {author} {\bibfnamefont {B.}~\bibnamefont
  {H{\"o}ckendorf}}, \bibinfo {author} {\bibfnamefont {A.}~\bibnamefont
  {Alvermann}}, \ and\ \bibinfo {author} {\bibfnamefont {H.}~\bibnamefont
  {Fehske}},\ }\href@noop {} {\bibfield  {journal} {\bibinfo  {journal} {arXiv
  preprint arXiv:1911.11413}\ } (\bibinfo {year} {2019})}\BibitemShut {NoStop}%
\bibitem [{\citenamefont {Rodr{\'\i}guez-Mena}\ and\ \citenamefont
  {Torres}(2019)}]{rodriguez2019topological}%
  \BibitemOpen
  \bibfield  {author} {\bibinfo {author} {\bibfnamefont {E.~A.}\ \bibnamefont
  {Rodr{\'\i}guez-Mena}}\ and\ \bibinfo {author} {\bibfnamefont {L.~F.}\
  \bibnamefont {Torres}},\ }\href@noop {} {\bibfield  {journal} {\bibinfo
  {journal} {Physical Review B}\ }\textbf {\bibinfo {volume} {100}},\ \bibinfo
  {pages} {195429} (\bibinfo {year} {2019})}\BibitemShut {NoStop}%
\bibitem [{\citenamefont {H{\"o}ckendorf}\ \emph {et~al.}(2020)\citenamefont
  {H{\"o}ckendorf}, \citenamefont {Alvermann},\ and\ \citenamefont
  {Fehske}}]{hockendorf2020cutting}%
  \BibitemOpen
  \bibfield  {author} {\bibinfo {author} {\bibfnamefont {B.}~\bibnamefont
  {H{\"o}ckendorf}}, \bibinfo {author} {\bibfnamefont {A.}~\bibnamefont
  {Alvermann}}, \ and\ \bibinfo {author} {\bibfnamefont {H.}~\bibnamefont
  {Fehske}},\ }\href@noop {} {\bibfield  {journal} {\bibinfo  {journal} {arXiv
  preprint arXiv:2004.03290}\ } (\bibinfo {year} {2020})}\BibitemShut {NoStop}%
\bibitem [{\citenamefont {Alvarez}\ \emph
  {et~al.}(2018{\natexlab{b}})\citenamefont {Alvarez}, \citenamefont {Vargas},
  \citenamefont {Berdakin},\ and\ \citenamefont
  {Torres}}]{alvarez2018topological}%
  \BibitemOpen
  \bibfield  {author} {\bibinfo {author} {\bibfnamefont {V.~M.}\ \bibnamefont
  {Alvarez}}, \bibinfo {author} {\bibfnamefont {J.~B.}\ \bibnamefont {Vargas}},
  \bibinfo {author} {\bibfnamefont {M.}~\bibnamefont {Berdakin}}, \ and\
  \bibinfo {author} {\bibfnamefont {L.~F.}\ \bibnamefont {Torres}},\
  }\href@noop {} {\bibfield  {journal} {\bibinfo  {journal} {The European
  Physical Journal Special Topics}\ }\textbf {\bibinfo {volume} {227}},\
  \bibinfo {pages} {1295} (\bibinfo {year} {2018}{\natexlab{b}})}\BibitemShut
  {NoStop}%
\bibitem [{\citenamefont {Fang}\ \emph {et~al.}(2016)\citenamefont {Fang},
  \citenamefont {Weng}, \citenamefont {Dai},\ and\ \citenamefont
  {Fang}}]{fang2016topological}%
  \BibitemOpen
  \bibfield  {author} {\bibinfo {author} {\bibfnamefont {C.}~\bibnamefont
  {Fang}}, \bibinfo {author} {\bibfnamefont {H.}~\bibnamefont {Weng}}, \bibinfo
  {author} {\bibfnamefont {X.}~\bibnamefont {Dai}}, \ and\ \bibinfo {author}
  {\bibfnamefont {Z.}~\bibnamefont {Fang}},\ }\href@noop {} {\bibfield
  {journal} {\bibinfo  {journal} {Chinese Physics B}\ }\textbf {\bibinfo
  {volume} {25}},\ \bibinfo {pages} {117106} (\bibinfo {year}
  {2016})}\BibitemShut {NoStop}%
\bibitem [{\citenamefont {Diehl}\ \emph {et~al.}(2011)\citenamefont {Diehl},
  \citenamefont {Rico}, \citenamefont {Baranov},\ and\ \citenamefont
  {Zoller}}]{diehl2011topology}%
  \BibitemOpen
  \bibfield  {author} {\bibinfo {author} {\bibfnamefont {S.}~\bibnamefont
  {Diehl}}, \bibinfo {author} {\bibfnamefont {E.}~\bibnamefont {Rico}},
  \bibinfo {author} {\bibfnamefont {M.~A.}\ \bibnamefont {Baranov}}, \ and\
  \bibinfo {author} {\bibfnamefont {P.}~\bibnamefont {Zoller}},\ }\href@noop {}
  {\bibfield  {journal} {\bibinfo  {journal} {Nature Physics}\ }\textbf
  {\bibinfo {volume} {7}},\ \bibinfo {pages} {971} (\bibinfo {year}
  {2011})}\BibitemShut {NoStop}%
\bibitem [{\citenamefont {Shen}\ \emph {et~al.}(2018)\citenamefont {Shen},
  \citenamefont {Zhen},\ and\ \citenamefont {Fu}}]{shen2018topological}%
  \BibitemOpen
  \bibfield  {author} {\bibinfo {author} {\bibfnamefont {H.}~\bibnamefont
  {Shen}}, \bibinfo {author} {\bibfnamefont {B.}~\bibnamefont {Zhen}}, \ and\
  \bibinfo {author} {\bibfnamefont {L.}~\bibnamefont {Fu}},\ }\href@noop {}
  {\bibfield  {journal} {\bibinfo  {journal} {Physical review letters}\
  }\textbf {\bibinfo {volume} {120}},\ \bibinfo {pages} {146402} (\bibinfo
  {year} {2018})}\BibitemShut {NoStop}%
\bibitem [{\citenamefont {Usaj}\ \emph {et~al.}(2014)\citenamefont {Usaj},
  \citenamefont {Perez-Piskunow}, \citenamefont {Torres},\ and\ \citenamefont
  {Balseiro}}]{usaj2014irradiated}%
  \BibitemOpen
  \bibfield  {author} {\bibinfo {author} {\bibfnamefont {G.}~\bibnamefont
  {Usaj}}, \bibinfo {author} {\bibfnamefont {P.~M.}\ \bibnamefont
  {Perez-Piskunow}}, \bibinfo {author} {\bibfnamefont {L.~F.}\ \bibnamefont
  {Torres}}, \ and\ \bibinfo {author} {\bibfnamefont {C.~A.}\ \bibnamefont
  {Balseiro}},\ }\href@noop {} {\bibfield  {journal} {\bibinfo  {journal}
  {Physical Review B}\ }\textbf {\bibinfo {volume} {90}},\ \bibinfo {pages}
  {115423} (\bibinfo {year} {2014})}\BibitemShut {NoStop}%
\bibitem [{\citenamefont {Wang}\ \emph {et~al.}(2012)\citenamefont {Wang},
  \citenamefont {Sun}, \citenamefont {Chen}, \citenamefont {Franchini},
  \citenamefont {Xu}, \citenamefont {Weng}, \citenamefont {Dai},\ and\
  \citenamefont {Fang}}]{wang2012dirac}%
  \BibitemOpen
  \bibfield  {author} {\bibinfo {author} {\bibfnamefont {Z.}~\bibnamefont
  {Wang}}, \bibinfo {author} {\bibfnamefont {Y.}~\bibnamefont {Sun}}, \bibinfo
  {author} {\bibfnamefont {X.-Q.}\ \bibnamefont {Chen}}, \bibinfo {author}
  {\bibfnamefont {C.}~\bibnamefont {Franchini}}, \bibinfo {author}
  {\bibfnamefont {G.}~\bibnamefont {Xu}}, \bibinfo {author} {\bibfnamefont
  {H.}~\bibnamefont {Weng}}, \bibinfo {author} {\bibfnamefont {X.}~\bibnamefont
  {Dai}}, \ and\ \bibinfo {author} {\bibfnamefont {Z.}~\bibnamefont {Fang}},\
  }\href@noop {} {\bibfield  {journal} {\bibinfo  {journal} {Physical Review
  B}\ }\textbf {\bibinfo {volume} {85}},\ \bibinfo {pages} {195320} (\bibinfo
  {year} {2012})}\BibitemShut {NoStop}%
\bibitem [{\citenamefont {Wang}\ \emph
  {et~al.}(2013{\natexlab{b}})\citenamefont {Wang}, \citenamefont {Weng},
  \citenamefont {Wu}, \citenamefont {Dai},\ and\ \citenamefont
  {Fang}}]{wang2013three}%
  \BibitemOpen
  \bibfield  {author} {\bibinfo {author} {\bibfnamefont {Z.}~\bibnamefont
  {Wang}}, \bibinfo {author} {\bibfnamefont {H.}~\bibnamefont {Weng}}, \bibinfo
  {author} {\bibfnamefont {Q.}~\bibnamefont {Wu}}, \bibinfo {author}
  {\bibfnamefont {X.}~\bibnamefont {Dai}}, \ and\ \bibinfo {author}
  {\bibfnamefont {Z.}~\bibnamefont {Fang}},\ }\href@noop {} {\bibfield
  {journal} {\bibinfo  {journal} {Physical Review B}\ }\textbf {\bibinfo
  {volume} {88}},\ \bibinfo {pages} {125427} (\bibinfo {year}
  {2013}{\natexlab{b}})}\BibitemShut {NoStop}%
\bibitem [{\citenamefont {Midya}\ \emph {et~al.}(2018)\citenamefont {Midya},
  \citenamefont {Zhao},\ and\ \citenamefont {Feng}}]{midya2018non}%
  \BibitemOpen
  \bibfield  {author} {\bibinfo {author} {\bibfnamefont {B.}~\bibnamefont
  {Midya}}, \bibinfo {author} {\bibfnamefont {H.}~\bibnamefont {Zhao}}, \ and\
  \bibinfo {author} {\bibfnamefont {L.}~\bibnamefont {Feng}},\ }\href@noop {}
  {\bibfield  {journal} {\bibinfo  {journal} {Nature communications}\ }\textbf
  {\bibinfo {volume} {9}},\ \bibinfo {pages} {1} (\bibinfo {year}
  {2018})}\BibitemShut {NoStop}%
\bibitem [{\citenamefont {Dembowski}\ \emph {et~al.}(2001)\citenamefont
  {Dembowski}, \citenamefont {Gr{\"a}f}, \citenamefont {Harney}, \citenamefont
  {Heine}, \citenamefont {Heiss}, \citenamefont {Rehfeld},\ and\ \citenamefont
  {Richter}}]{dembowski2001experimental}%
  \BibitemOpen
  \bibfield  {author} {\bibinfo {author} {\bibfnamefont {C.}~\bibnamefont
  {Dembowski}}, \bibinfo {author} {\bibfnamefont {H.-D.}\ \bibnamefont
  {Gr{\"a}f}}, \bibinfo {author} {\bibfnamefont {H.}~\bibnamefont {Harney}},
  \bibinfo {author} {\bibfnamefont {A.}~\bibnamefont {Heine}}, \bibinfo
  {author} {\bibfnamefont {W.}~\bibnamefont {Heiss}}, \bibinfo {author}
  {\bibfnamefont {H.}~\bibnamefont {Rehfeld}}, \ and\ \bibinfo {author}
  {\bibfnamefont {A.}~\bibnamefont {Richter}},\ }\href@noop {} {\bibfield
  {journal} {\bibinfo  {journal} {Physical review letters}\ }\textbf {\bibinfo
  {volume} {86}},\ \bibinfo {pages} {787} (\bibinfo {year} {2001})}\BibitemShut
  {NoStop}%
\bibitem [{\citenamefont {Ding}\ \emph {et~al.}(2016)\citenamefont {Ding},
  \citenamefont {Ma}, \citenamefont {Xiao}, \citenamefont {Zhang},\ and\
  \citenamefont {Chan}}]{ding2016emergence}%
  \BibitemOpen
  \bibfield  {author} {\bibinfo {author} {\bibfnamefont {K.}~\bibnamefont
  {Ding}}, \bibinfo {author} {\bibfnamefont {G.}~\bibnamefont {Ma}}, \bibinfo
  {author} {\bibfnamefont {M.}~\bibnamefont {Xiao}}, \bibinfo {author}
  {\bibfnamefont {Z.}~\bibnamefont {Zhang}}, \ and\ \bibinfo {author}
  {\bibfnamefont {C.~T.}\ \bibnamefont {Chan}},\ }\href@noop {} {\bibfield
  {journal} {\bibinfo  {journal} {Physical Review X}\ }\textbf {\bibinfo
  {volume} {6}},\ \bibinfo {pages} {021007} (\bibinfo {year}
  {2016})}\BibitemShut {NoStop}%
\bibitem [{\citenamefont {Cerjan}\ \emph {et~al.}(2019)\citenamefont {Cerjan},
  \citenamefont {Huang}, \citenamefont {Wang}, \citenamefont {Chen},
  \citenamefont {Chong},\ and\ \citenamefont
  {Rechtsman}}]{cerjan2019experimental}%
  \BibitemOpen
  \bibfield  {author} {\bibinfo {author} {\bibfnamefont {A.}~\bibnamefont
  {Cerjan}}, \bibinfo {author} {\bibfnamefont {S.}~\bibnamefont {Huang}},
  \bibinfo {author} {\bibfnamefont {M.}~\bibnamefont {Wang}}, \bibinfo {author}
  {\bibfnamefont {K.~P.}\ \bibnamefont {Chen}}, \bibinfo {author}
  {\bibfnamefont {Y.}~\bibnamefont {Chong}}, \ and\ \bibinfo {author}
  {\bibfnamefont {M.~C.}\ \bibnamefont {Rechtsman}},\ }\href@noop {} {\bibfield
   {journal} {\bibinfo  {journal} {Nature Photonics}\ }\textbf {\bibinfo
  {volume} {13}},\ \bibinfo {pages} {623} (\bibinfo {year} {2019})}\BibitemShut
  {NoStop}%
\bibitem [{\citenamefont {Jotzu}\ \emph {et~al.}(2014)\citenamefont {Jotzu},
  \citenamefont {Messer}, \citenamefont {Desbuquois}, \citenamefont {Lebrat},
  \citenamefont {Uehlinger}, \citenamefont {Greif},\ and\ \citenamefont
  {Esslinger}}]{jotzu2014experimental}%
  \BibitemOpen
  \bibfield  {author} {\bibinfo {author} {\bibfnamefont {G.}~\bibnamefont
  {Jotzu}}, \bibinfo {author} {\bibfnamefont {M.}~\bibnamefont {Messer}},
  \bibinfo {author} {\bibfnamefont {R.}~\bibnamefont {Desbuquois}}, \bibinfo
  {author} {\bibfnamefont {M.}~\bibnamefont {Lebrat}}, \bibinfo {author}
  {\bibfnamefont {T.}~\bibnamefont {Uehlinger}}, \bibinfo {author}
  {\bibfnamefont {D.}~\bibnamefont {Greif}}, \ and\ \bibinfo {author}
  {\bibfnamefont {T.}~\bibnamefont {Esslinger}},\ }\href@noop {} {\bibfield
  {journal} {\bibinfo  {journal} {Nature}\ }\textbf {\bibinfo {volume} {515}},\
  \bibinfo {pages} {237} (\bibinfo {year} {2014})}\BibitemShut {NoStop}%
\bibitem [{\citenamefont {Li}\ \emph {et~al.}(2019)\citenamefont {Li},
  \citenamefont {Lee},\ and\ \citenamefont {Gong}}]{li2019topology}%
  \BibitemOpen
  \bibfield  {author} {\bibinfo {author} {\bibfnamefont {L.}~\bibnamefont
  {Li}}, \bibinfo {author} {\bibfnamefont {C.~H.}\ \bibnamefont {Lee}}, \ and\
  \bibinfo {author} {\bibfnamefont {J.}~\bibnamefont {Gong}},\ }\href@noop {}
  {\bibfield  {journal} {\bibinfo  {journal} {arXiv preprint arXiv:1910.03229}\
  } (\bibinfo {year} {2019})}\BibitemShut {NoStop}%
\bibitem [{\citenamefont {St-Jean}\ \emph {et~al.}(2017)\citenamefont
  {St-Jean}, \citenamefont {Goblot}, \citenamefont {Galopin}, \citenamefont
  {Lema{\^\i}tre}, \citenamefont {Ozawa}, \citenamefont {Le~Gratiet},
  \citenamefont {Sagnes}, \citenamefont {Bloch},\ and\ \citenamefont
  {Amo}}]{st2017lasing}%
  \BibitemOpen
  \bibfield  {author} {\bibinfo {author} {\bibfnamefont {P.}~\bibnamefont
  {St-Jean}}, \bibinfo {author} {\bibfnamefont {V.}~\bibnamefont {Goblot}},
  \bibinfo {author} {\bibfnamefont {E.}~\bibnamefont {Galopin}}, \bibinfo
  {author} {\bibfnamefont {A.}~\bibnamefont {Lema{\^\i}tre}}, \bibinfo {author}
  {\bibfnamefont {T.}~\bibnamefont {Ozawa}}, \bibinfo {author} {\bibfnamefont
  {L.}~\bibnamefont {Le~Gratiet}}, \bibinfo {author} {\bibfnamefont
  {I.}~\bibnamefont {Sagnes}}, \bibinfo {author} {\bibfnamefont
  {J.}~\bibnamefont {Bloch}}, \ and\ \bibinfo {author} {\bibfnamefont
  {A.}~\bibnamefont {Amo}},\ }\href@noop {} {\bibfield  {journal} {\bibinfo
  {journal} {Nature Photonics}\ }\textbf {\bibinfo {volume} {11}},\ \bibinfo
  {pages} {651} (\bibinfo {year} {2017})}\BibitemShut {NoStop}%
\bibitem [{\citenamefont {Bergholtz}\ and\ \citenamefont
  {Budich}(2019)}]{bergholtz2019non}%
  \BibitemOpen
  \bibfield  {author} {\bibinfo {author} {\bibfnamefont {E.~J.}\ \bibnamefont
  {Bergholtz}}\ and\ \bibinfo {author} {\bibfnamefont {J.~C.}\ \bibnamefont
  {Budich}},\ }\href@noop {} {\bibfield  {journal} {\bibinfo  {journal}
  {Physical Review Research}\ }\textbf {\bibinfo {volume} {1}},\ \bibinfo
  {pages} {012003} (\bibinfo {year} {2019})}\BibitemShut {NoStop}%
\end{thebibliography}

%
\end{document}